\documentclass[preprint,12pt]{elsarticle}

\PassOptionsToPackage{hyphens}{url}\usepackage[pdftex,colorlinks]{hyperref}
\hypersetup{
	colorlinks,
	linkcolor={red!50!blue},
	citecolor={blue!50!green},
	urlcolor={blue!80!black}
}

\graphicspath{{Figures/}}

\usepackage{longtable}
\usepackage{fontenc}
\usepackage{array}
\newcolumntype{P}[1]{>{\raggedright\arraybackslash}p{#1}}

\usepackage{amssymb}
\usepackage{amsmath}
\usepackage{amsthm}
\usepackage{color}

\usepackage{caption}
\usepackage{subcaption}

\newcommand{\fig}[1]{\text{Fig.\ \ref{#1}}}
\newcommand{\eq}[1]{\text{Eq.\ (\ref{#1})}}
\newcommand{\tbl}[1]{\text{Table \ref{#1}}}
\newcommand{\sect}[1]{\text{Section \ref{#1}}}

\definecolor{paper_blue}{rgb}{0.3,0.2,0.75}
\definecolor{paper_red}{rgb}{0.65,0.1,0.15}
\definecolor{paper_green}{rgb}{0.05,0.35,0.125}
\definecolor{paper_grey}{gray}{0.375}
\definecolor{perm}{rgb}{0.1,0.45,0.85}
\definecolor{deemph}{rgb}{0.7,0.7,0.7}

\usepackage{etoolbox}
\usepackage{tikz}

\usepackage{mmacells}

\definecolor{color_variable}{rgb}{0.1,0.55,0.25}
\definecolor{color_function}{rgb}{0.1,0.35,0.75}

\newcommand{\mmaFuncName}[1]{\text{\tt{\color{color_function}#1}}}
\makeatletter
\newcommand{\mmaFunc}[2]{\def\funcdata{}\foreach \data in {#2} {\protected@xappto\funcdata{{\tt[\,\data\,]}\!}}\mmaFuncName{#1}\funcdata\,}
\makeatother
\newcommand{\mmaFuncDef}[2]{\mmaFunc{#1}{#2}{\tt:}}
\newcommand{\mmaVar}[1]{\text{\tt{\color{color_variable}{\sl#1}}}}
\newcommand{\mmaVarDef}[1]{\mmaVar{#1\rule[-1.05pt]{7.5pt}{.75pt}}}

\newcommand{\mmaHashSymbol}{{\color{color_variable}\#}}

\newcommand{\Mat}{{M}}
\newcommand{\MatPositive}{{\Mat}_{+}}
\newcommand{\Le}{\textup{\protect\scalebox{-1}[1]{L}}}

\journal{}

\newcommand{\Mathematica}{{\textsc{Mathematica}~}}

\journal{Computer Physics Communications}

\begin{document}

\begin{frontmatter}



\title{Boundaries of the Amplituhedron with {\tt amplituhedronBoundaries}}

\author[TL]{Tomasz \L ukowski}
\author[RM]{Robert Moerman}
\address[TL]{t.lukowski@herts.ac.uk}
\address[RM]{r.moerman@herts.ac.uk}

\address{University of Hertfordshire, School of Physics, Astronomy and Mathematics, \\
AL10 9AB Hatfield,
United Kingdom.}

\begin{abstract}
Positive geometries provide a modern approach for computing scattering amplitudes in a variety of physical models. In order to facilitate the exploration of these new geometric methods, we introduce a \Mathematica package called ``{\tt amplituhedronBoundaries}'' for calculating the boundary structures of three positive geometries: the amplituhedron $\mathcal{A}_{n,k}^{(m)}$, the momentum amplituhedron $\mathcal{M}_{n,k}^{(m)}$ and the hypersimplex $\Delta_{k,n}$. The first two geometries are relevant for scattering amplitudes in planar $\mathcal{N}=4$ SYM, while the last one is a well-studied polytope appearing in many contexts in mathematics, and is closely related to $\mathcal{M}_{n,k}^{(2)}$. The package includes an array of useful tools for the study of these positive geometries, including their boundary stratifications, drawing their boundary posets, and additional tools for manipulating combinatorial structures useful for positive Grassmannians.  \end{abstract}


\end{frontmatter}

\clearpage
\section*{Program Summary}
\noindent
{\em Program title:} {\tt amplituhedronBoundaries}\\[0.5em]
{\em Permanent link to code:} \\\url{https://github.com/mrmrob003/amplituhedronBoundaries} \\[0.5em]
{\em Licensing provisions:} GNU General Public License 3 (GPLv3)\\[0.5em]
{\em Programming language:} Wolfram \Mathematica 11.0 \\[0.5em]
{\em Operating system:} Tested on Linux and Mac OS X. \\[0.5em]
{\em Nature of problem:} The package facilitates the determination and study of the boundary stratifications for three positive geometries: the amplituhedron, the momentum amplituhedron, and the hypersimplex. The first two geometries are relevant for scattering amplitudes in planar $\mathcal{N}=4$ SYM, while the last one is a well-studied polytope appearing in many important contexts in mathematics.\\[0.5em]
{\em Solution method:} The package includes an array of useful tools for exploring the three aforementioned positive geometries, including their boundary stratifications, drawing their boundary posets, and additional tools for manipulating combinatorial structures useful for positive Grassmannians.\\[0.5em]
{\em Restrictions:} Wolfram \Mathematica 11.0 or above

\clearpage
\tableofcontents

\clearpage

\section{Introduction}
\label{Sec:intro}

Recent years have seen an explosion of new ideas in scattering amplitudes in theories of scalar and gauge fields as well as in gravity. One important new direction has been the description of scattering processes in term of positive geometries \cite{Arkani-Hamed:2017tmz}. In this geometric approach, scattering amplitudes are encoded in canonical differential forms with the property that as one approaches any boundary of the positive geometry, the differential form behaves logarithmically. The archetypical example of a positive geometry is the amplituhedron \cite{Arkani-Hamed:2013jha} defined in the momentum twistor space \cite{Hodges:2009hk} for which the differential form encodes tree-level scattering amplitudes and integrands of loop-level amplitudes for planar $ \mathcal{N}=4$ super Yang-Mills (SYM) theory. More recently, a positive geometry for the same amplitudes has been defined directly in the spinor-helicity space --  the momentum amplituhedron \cite{Damgaard:2019ztj}. Both geometries can be seen as images of a positive Grassmannian through particular linear maps determined by positive matrices. Importantly, many properties of amplituhedra descend directly from the properties of positive Grassmannians. In particular, the fact that they admit a decomposition in terms cells of various dimensions parametrized by decorated permutations, and that this defines the boundary stratification for the geometry, is inherited directly from the cell decomposition of the positive Grassmannian. The details of these decompositions differ for all three objects and there is a significant amount of interesting on-going research trying to understand these boundary structures. 

One reason for which we are interested in finding these boundary stratifications is that they have profound physical significance: they describe all possible physical singularities of scattering amplitudes, including the collinear and soft limits. They are also interesting from a purely mathematical point of view: these are geometric spaces defined in very simple terms, analogous to the definition of polytopes, but they have a more complicated and interesting combinatorial structure. Their topological nature, in particular the question of whether they are homeomorphic to closed balls, is also mostly an open and interesting problem, which has been only recently solved for the positive Grassmannian $G_+(k,n)$ \cite{Galashin:2017onl} and the amplituhedron $\mathcal{A}_{n,k}^{(1)}$ \cite{Karp:2016uax}, see also \cite{Lukowski:2019kqi} for the study of the $m=2$ ampliuhedron.

In this paper we introduce a \Mathematica package  
\begin{center}``\texttt{amplituhedronBoundaries}''\end{center}
which addresses the problem of finding the boundary stratifications for the tree amplituhedron $\mathcal{A}_{n,k}^{(m)}$ and the momentum amplituhedron $\mathcal{M}_{n,k}^{(m)}$. Our main focus is on the case when $m=2$, however the package works well also beyond that and provides tools for studying the physical case when $m=4$. For $m=2$ it was shown in \cite{LPW} that the momentum amplituhedron $\mathcal{M}_{n,k}^{(2)}$ shares many properties with the hypersimplex $\Delta_{k,n}$, the latter can be seen as an image of the positive Grassmmanian under a moment map. For this reason, we include in our package functions allowing for the study of boundaries of the hypersimplex $\Delta_{k,n}$. 

Our package relies on Jacob Bourjaily's ``{\tt positroid}'' package \cite{Bourjaily:2012gy} for the positroid stratification of positive Grassmannians. In particular, the boundary stratification of positive Grassmannians, which is our starting point for finding the boundary stratification of the three previously introduced positive geometries, is implemented solely using {\tt positroid} functions. To extend this analysis to amplituhedra and the hypersimplex, we begin by providing functions for calculating the dimensions of images of each positroid cell in $G_{+}(k,n)$ into the amplituhedron, the momentum amplituhedron and the hypersimplex. Using this information and the known structure of facets (co-dimension-one boundaries) of the amplituhedron, the momentum amplituhedron and the hypersimplex, we are able to generate the complete boundary stratifications for these geometries. 

\section{Background}
\label{Sec:background}

We start by defining four geometric spaces which will be relevant for our discussion: the positive Grassmannian $G_+(k,n)$, the amplituhedron $\mathcal{A}_{n,k}^{(m)}$, the momentum amplituhedron $\mathcal{M}_{n,k}^{(m)}$ and the hypersimplex $\Delta_{k,n}$. We refer the reader to \cite{Arkani-Hamed:2013jha} and \cite{Damgaard:2019ztj} for comprehensive reviews on how scattering amplitudes can be extracted from amplituhedra. In this paper we will be  interested in the boundary structure of these geometries.


\subsection{Positive Grassmannian}
The Grassmanian $G(k,n)$ is the space of $k$-planes in $n$ dimensions,
which we can take to be a space of $k\times n$ matrices modulo $GL(k)$ transformations:
\begin{equation}
C=\left(\begin{tabular}{cccc}
$c_{11}$&$c_{12}$&$\ldots$&$c_{1n}$\\
$\vdots$&$\vdots$&$\ddots$&$\vdots$\\
$c_{k1}$&$c_{k2}$&$\ldots$&$c_{kn}$
\end{tabular}\right)
\end{equation}
The totally non-negative part of the Grassmannian $G_+(k,n)$ is a subset of the Grassmannian $G(k,n)$ consisting of elements described by matrices with all ordered maximal minors non-negative. Abusing notation, we will often refer to $G_+(k,n)$ as the \emph{positive Grassmannian}. The positive Grassmannian has been studied by Postnikov \cite{Postnikov:2006kva} and is known to have a very rich and interesting combinatorial structure. Each boundary stratum of $G_+(k,n)$ is called a \emph{positroid cell} and can be labelled by a variety of combinatorial objects, including decorated permutations, (equivalence classes of) plabic diagrams and $\Le$-diagrams. A {\it decorated permutation} is a generalization of ordinary permutation which allows for two types of fixed-points. It is a map $\pi:\{1,2,\ldots,n\}\to\{1,2,\ldots,2n\}$ such that $a\leq \pi(a)\leq a+n$. We will use decorated permutations to label positroid cells $S_\pi\subset G_+(k,n)$, but also to label images of $S_\pi$ in the amplituhedron, the momentum amplituhedron and the hypersimplex. For a given positroid cell $S_\pi\subset G_+(k,n)$ we denote by $\dim_C \pi$ its dimension and by $\partial_C\pi$ its boundary stratification. Moreover, we denote by $\partial^{-1}_C\pi$ the inverse boundary stratification of $\pi$, i.e.\ the set of all positroid cells $S_{\pi'}\subset G_+(k,n)$ for which $\pi\in\partial_C\pi'$.



\subsection{Amplituhedron}
The tree amplituhedron $\mathcal{A}_{n,k}^{(m)}$ encodes tree-level scattering amplitudes in planar $\mathcal{N}=4$ SYM \cite{Arkani-Hamed:2013jha}. It is defined as the image of the positive Grassmannian $G_+(k,n)$ through the map
\begin{equation}
\Phi_Z:G_{+}(k,n)\to G(k,k+m)\,,
\end{equation}
induced by a positive $(k+m)\times n$ matrix $Z\in\MatPositive(k+m,n)$, where $\MatPositive(k+m,n)$ is the space of all $(k+m)\times n$ matrices  with all maximal ordered minors positive. For an element of the positive Grassmannian $C=\{c_{\alpha i}\}\in G_{+}(k,n)$, the map $\Phi_Z$ is defined as
\begin{equation}
Y_\alpha^A= \Phi_Z(C)=c_{\alpha i}Z_i^A\,.
\end{equation}

One of the most important functions of our package is used is to calculate the dimension of the image of a given positroid cell  through the map $\Phi_Z$. For $S_\pi\subset G_+(k,n)$ we define
\begin{equation}
\dim_A\pi=\dim \Phi_Z(S_\pi).
\end{equation}
In order to calculate $\dim_A\pi$ we need to find the dimension of the tangent space to $\Phi_Z(C_\pi)$. This can be accomplished by parametrizing the cell $S_\pi$ using canonical coordinates $\beta_i$ and choosing a generic point $P$ in $S_\pi$. The dimension $\dim_A\pi$ can then be computed as the rank of the matrix
\begin{equation}
\frac{\partial \Phi_Z(C_\pi(P))_{\alpha}^A}{\partial \beta_i},\qquad
\alpha=1,\ldots,k,\,\, A=1,\ldots,k+m,\,\,\\
i=1,\ldots,\dim_C\pi\,.
\end{equation}

It is always true that $\dim_C(\pi)\geq \dim_A(\pi)$ and we can distinguish two cases:
\begin{itemize}
\item $\dim_C(\pi)=\dim_A(\pi)$: we will refer to cells satisfying this condition as simplicial-like,
\item $\dim_C(\pi)>\dim_A(\pi)$: these cells are polytopal-like.
\end{itemize}
This distinction refers to properties of polytopes: a simplex is a polytope which cannot be subdivided into smaller polytopes without introducing new vertices. Similarly here, the simplicial-like cells are those for which their images cannot be subdivided into smaller images of positroid cells. This is not the case for polytopal-like cells. In particular, for a positroid cell $\pi$ for which $\dim_C(\pi)>\dim_A(\pi)$, we can find a collection of cells in its boundary stratification $\partial_C\pi$, with the same amplituhedron dimension as $\pi$. Moreover, there exists a (non-unique) subset $\{\pi_1,\ldots,\pi_r\}\in\partial_C\pi$ such that the images $\{\Phi_Z(\pi_1),\ldots,\Phi_Z(\pi_r)\}$ triangulate the image $\Phi_Z(\pi)$. 
For $n>m+k$, the dimension of $G_+(k,n)$ is larger than the dimension of $G(k,k+m)$, i.e.~the amplituhedron $\mathcal{A}_{n,k}^{(m)}$ itself is polytopal-like, and therefore the map $\Phi_Z$ is not injective. In order to have an injective map, we need to find a collection of $(m\cdot k)$-dimensional positroid cells $\mathcal{C}=\{\pi_1,\ldots,\pi_p\}$ such that 
\begin{itemize}
	\item they are mapped injectively to the amplituhedron,
	\item their images are disjoint, and
	\item the union of these images is dense in $\mathcal{A}_{n,k}^{(m)}$.
\end{itemize}
We say that such a collection triangulates the amplituhedron, and we call the cells $\pi_i\in\mathcal{C}$ (or equivalently their images) the generalized triangles. If additionally the co-dimension-one boundaries of the images are compatible, then we say that the collection $\mathcal{C}$ is a {\it good triangulation} of the amplituhedron, see \cite{LPW} for more details. It has been conjectured that it is always possible to find a triangulation for a given amplituhedron and candidate triangulations are given by the BCFW recursion relation, see e.g.\ \cite{Karp:2017ouj}. For $m=1$ and $m=2$ this conjecture has already been proven in \cite{Karp:2016uax} and \cite{Bao:2019bfe}, respectively.

We can now turn to studying the boundaries of the amplituhedron. The co-dimension-one boundaries of $\mathcal{A}_{n,k}^{(m)}$ are known for small values of $m$, see e.g.\ \cite{Arkani-Hamed:2017tmz}. If we define a $SL(m+k)$ invariant bracket
\begin{equation}\label{inv.bracket}
\langle W_{a_1}\ldots W_{a_{k+m}}\rangle=\epsilon_{A_1A_2\ldots A_{k+m}}W^{A_1}_{a_1}W^{A_2}_{a_2}\ldots W^{A_{k+m}}_{a_{k+m}}\,,
\end{equation}
where $W_{a_j}$ can be one of $Y_\alpha$ or $Z_i$, then the co-dimension-one boundaries are
\begin{itemize}
\item for $m=1$:  $\langle Yi\rangle\equiv\langle YZ_i\rangle=0$, for $i=1,\ldots,n$,
\item for $m=2$:  $\langle Yii+1\rangle\equiv\langle YZ_iZ_{i+1}\rangle=0$, for $i=1,\ldots,n$,
\item for $m=4$:  $\langle Yii+1jj+1\rangle\equiv\langle YZ_iZ_{i+1}Z_jZ_{j+1}\rangle=0$, for $i<j=1,\ldots,n$.
\end{itemize}
Here $Y=Y_1\ldots Y_k$, where $Y_i$ denotes the $i^\text{th}$ column of $Y$.
Starting from co-dimension-one boundaries we can find all lower dimensional ones following a simple recursive procedure \cite{Lukowski:2019kqi}: assume that we have found all amplituhedron boundaries of amplituhedron dimension larger than $d$. Let us study all positroid cells $\pi$ with amplituhedron dimension $\dim_A \pi = d$. For a given cell $\pi$, there are two options:
\begin{itemize}
\item either the amplituhedron dimension for all inverse boundaries of $\pi$ are higher than the amplituhedron dimension of $\pi$: $\forall_{\pi'\in\partial^{-1}\pi}: \dim_A\pi'>\dim_A\pi$;
\item or we can find a cell among the inverse boundaries of $\pi$ which has a higher Grassmannian dimension but the same amplituhedron dimension as $\pi$: $\exists_{\pi'\in\partial^{-1}\pi}:\dim_A\pi'=\dim_A\pi\text{ and }\dim_C\pi'>\dim_C\pi$.
\end{itemize}
We only keep the former cells, which we will refer to as \emph{faces}, since the latter are necessarily elements of a triangulation of a boundary of the amplituhedron. After discarding these latter cells, there is still a possibility that some of the remaining cell images are spurious boundaries, which arise as spurious faces in triangulations of polytopal-like boundaries.
Spurious boundaries can be identified (and removed) because they belong to a single $(d+1)$-dimensional amplituhedron boundary, while external boundaries belong to at least two such boundaries. This procedure allows us to find all external boundaries of dimension $d$. We can follow this procedure recursively, starting from the known co-dimension-one boundaries, and work our way down to zero-dimensional boundaries: points. 


\subsection{Momentum Amplituhedron}
The tree momentum amplituhedron $\mathcal{M}_{n,k}^{(m)}$ has been introduced in \cite{Damgaard:2019ztj} for $m=4$ and generalized to any even $m$ in \cite{LPW}. It is a positive geometry encoding tree scattering amplitudes in planar $\mathcal{N}=4$ SYM  directly in the spinor-helicity space. Similar to the definition of the ordinary amplituhedron, the momentum amplituhedron is defined as the image of the positive Grassmannian $G_+(k,n)$ through the map
\begin{equation}
\label{Phi}
\Phi_{(\Lambda,\tilde\Lambda)}:G_+(k,n)\to G(k,k+\tfrac{m}{2})\times G(n-k,n-k+\tfrac{m}{2})\,,
\end{equation}
Here $\tilde \Lambda$ is a $(k+\frac{m}{2})\times n$ positive matrix and $\Lambda$ is a $(n-k+\frac{m}{2})\times n$ matrix for which the orthogonal complement $\Lambda^\perp$ is a positive matrix.
To each element $C=\{c_{\dot\alpha i}\}$ of the positive Grassmannian $G_{+}(k,n)$ the map $\Phi_{\Lambda,\tilde\Lambda}$ associates a pair of Grassmannian elements $(\tilde Y,Y)\in G(k,k+\tfrac{m}{2})\times G(n-k,n-k+\tfrac{m}{2})$ in the following way
\begin{align}
\tilde Y^{\dot{A}}_{\dot{\alpha}}=c_{\dot\alpha i}\,\tilde\Lambda_i^{\dot{A}}\,,\qquad Y^A_\alpha=c^\perp_{\alpha i}\,\Lambda_i^A\,,
\label{Y}
\end{align}
where $C^\perp=\{c^\perp_{\alpha i}\}$ is the orthogonal complement of $C$. One can show that $Y$ has rank $(n-k)$, and therefore it is an element of $G(n-k,n-k+\tfrac{m}{2})$ and the map $\Phi_{(\Lambda,\tilde\Lambda)}$ is well defined.

Although the dimension of the $(Y,\tilde Y)$-space is $\frac{k\cdot m}{2}+(n-k)\frac{m}{2}=\frac{n\cdot m}{2}$, the image of the positive Grassmannian through the $\Phi_{(\Lambda,\tilde\Lambda)}$ map is $\frac{m}{2}(n-\frac{m}{2})$-dimensional since one can show that the following relation holds true
\begin{equation}
(Y^\perp\Lambda)\cdot(\tilde Y^\perp \tilde\Lambda)=0,
\end{equation}  
and therefore the image is embedded in a surface of co-dimension $\left(\frac{m}{2}\right)^2$.

Facets of the momentum amplituhedron $\mathcal{M}_{n,k}^{(4)}$ have been studied in \cite{Damgaard:2019ztj} and they belong to one of the following classes:
\begin{equation}
\langle Y i i+1\rangle=0\,,\qquad [\tilde Y ii+1]=0\,\qquad S_{i,i+1,\ldots,j}=0\,,
\end{equation}
where 
\begin{equation}
S_{i,i+1,\ldots j}=\sum\limits_{a<b=i}^j \langle Y a b\rangle [\tilde Y ab]\,,
\end{equation}
are equivalent to the Mandelstam invariants written in the momentum amplituhedron space. Here we defined invariant brackets $\langle \rangle$ and $[]$ analogously to \eq{inv.bracket}.

The facets of $\mathcal{M}_{n,k}^{(2)}$ have the form
\begin{equation}
\langle Yi\rangle=0\,,\qquad [\tilde Y i]=0\,.
\end{equation}
In \cite{LPW} it was conjectured that the momentum amplituhedron $\mathcal{M}_{n,k}^{(2)}$ is closely related to the hypersimplex $\Delta_{k,n}$ which we define in the next section. In particular, many combinatorial properties  for these two objects are identical, e.g. they have the same dissections and triangulations. Our package allows one to check that they have identical boundary stratifications.

\subsection{Hypersimplex}
The hypersimplex $\Delta_{k,n}$ is defined as the convex hull of the points $e_I=\sum_{i\in I}e_i$ where $I$ is a $k$-element subset of $[n]\equiv\{1,2,\ldots,n\}$ and $e_i$ are standard basis vectors in $\mathbb{R}^n$. Since for all $x=(x_1,\ldots,x_n)\in\Delta_{k,n}$ we have $x_1+\ldots+x_n=k$ then the hypersimplex $\Delta_{k,n}$ is $(n-1)$-dimensional.

Equivalently, the hypersimplex $\Delta_{k,n}$ is the image of the positive Grassmannian $G_+(k,n)$ through the moment map
$$
\mu:G_+(k,n)\to\mathbb{R}^n\,,
$$
defined as
$$
\mu(C)=\frac{\sum_I p_I(C)e_I}{\sum_I p_I(C)}\,,
$$
where $C\in G_+(k,n)$ is an element of the positive Grassmannian $G_{+}(k,n)$ and $p_I(C)$ is the $I^\text{th}$ Pl\"{u}cker variable, i.e.\ the maximal minor formed of columns of $C$ labelled by elements of $I$.

The hypersimplex $\Delta_{k,n}$ has exactly $2n$ facets: $n$ of them satisfying $x_i=0$ and $n$ of them satisfying $x_i=1$ for $i=1,\ldots,n$. The former are identical with $\Delta_{k-1,n-1}$ while the latter are identical with $\Delta_{k,n-1}$.

\section{Installation and Setup}
\label{sec:setup}

\subsection*{Dependencies}
\label{sec:setup-dependencies}

The {\tt amplituhedronBoundaries} package has the following \Mathematica package dependencies:
\begin{itemize}
	\item {\tt MaTeX}: for creating \LaTeX~labels in diagrams. This package can be installed following the instructions found on: \url{https://github.com/szhorvat/MaTeX}.
	\item {\tt positroids}: for finding the stratification of boundaries in the positive Grassmannian. This package was written by Jacob Bourjaily \cite{Bourjaily:2012gy} and it is available for download from his arXiv submission: \url{https://arxiv.org/e-print/1212.6974}.
\end{itemize}

\subsection*{Installation}
\label{sec:setup-install}
The {\tt amplituhedronBoundaries} package is available on GitHub\footnote{The package {\tt amplituhedronBoundaries} can be found at: \url{https://raw.githubusercontent.com/mrmrob003/amplituhedronBoundariesTest/master/amplituhedronBoundaries}.} and can be download using
{\small
	\begin{mmaCell}[index=1,moredefined={URLDownload}]{Input}
SetDirectory[NotebookDirectory[]];
URLDownload["https://raw.githubusercontent.com/mrmrob003/
	amplituhedronBoundaries/master/"<>\mmaHashSymbol,\mmaHashSymbol]\&@
	"amplituhedronBoundaries.m";
	\end{mmaCell}
}
\noindent which downloads the package into directory of the notebook used to execute the above command. 

Once the {\tt amplituhedronBoundaries} package has been downloaded, together with the pre-requisite packages listed above, it can be loaded using
{\small
	\begin{mmaCell}[moredefined={amplituhedronBoundaries}]{Input}
SetDirectory[NotebookDirectory[]];
<<amplituhedronBoundaries\`
	\end{mmaCell}
}
\noindent After loading the package
\begin{enumerate}[(i)]
	\item a local data directory called {\tt Data/} is created in the same directory of the notebook, and
	\item a private variable called {\tt \$cache} is set to {\tt False}.
\end{enumerate} 
Having {\tt \$cache=False} means that the outputs of all functions are stored in the kernel's memory only and not to file. However, one can choose to have the outputs of some functions saved to file by setting {\tt \$cache=True} before loading the package: e.g.\
{\small
	\begin{mmaCell}[moredefined={amplituhedronBoundaries,cache}]{Input}
SetDirectory[NotebookDirectory[]];
\$cache=True;
<<amplituhedronBoundaries\`
	\end{mmaCell}
}
\noindent By setting {\tt \$cache=True} the outputs of the following list of functions are stored to file in the local data directory {\tt Data/}:
\begin{itemize}
	\item {\tt ampDimension}	
	\item {\tt ampStratification}	
	\item {\tt ampGeneralizedTriangles}
	\item {\tt momDimension}	
	\item {\tt momStratification}
	\item {\tt hypBases}
	\item {\tt hypDimension}	
	\item {\tt hypStratification}
\end{itemize}
Storing these outputs to file has the advantage that they will never need to be explicitly calculated again as they can be looked-up  almost immediately by the package. The disadvantage of choosing to store outputs to file is that the first-time evaluation of the above functions is longer.

\subsection*{Prepared Data Files}
A collection of data files containing the stored outputs of the functions listed previously evaluated for a number of examples has been prepared and is available on GitHub. These prepared files can be downloaded and added to the local data directory {\tt Data/} by executing the following command:
{\small
	\begin{mmaCell}[moredefined={fetchData}]{Input}
fetchData[]
	\end{mmaCell}
}
\noindent More details on the {\tt fetchData} function are available in \sect{sec:soft-aux}.

\section{Software}
\label{sec:software}
Our package is divided into four classes of functions, which can be distinguished by their unique name prefix:
\begin{itemize}
\item \emph{pos} -- positive Grassmannian $G_+(k,n)$
\item \emph{amp} -- amplituhedron $\mathcal{A}_{n,k}^{(m)}$
\item \emph{mom} -- momentum amplituhedron $\mathcal{M}_{n,k}^{(m)}$
\item \emph{hyp} -- hypersimplex $\Delta_{k,n}$
\end{itemize}

\subsection{Auxiliary Functions}\label{sec:soft-aux}
\begin{itemize}
\item \mmaFuncDef{fetchData}{\!} downloads a collection of prepared data files which can be used by {\tt amplituhedronBoundaries} to instantaneously look-up outputs that have already been computed. To this end, the function downloads the tarball {\tt Data.tar.gz} from GitHub\footnote{The tarball {\tt Data.tar.gz} can be found at: \url{https://raw.githubusercontent.com/mrmrob003/amplituhedronBoundaries/master/Data.tar.gz}.}, unzips the tarball and adds any data files to {\tt Data/} which are not already there; any data files which \emph{are} already present in {\tt Data/} will not be overwritten.
	
\item \mmaFuncDef{printCacheStatus}{\!} displays the value of the variable {\tt \$cache}. The default value of this variable is {\tt False}.
{\small
\begin{mmaCell}[index=1,moredefined={printCacheStatus}]{Input}
printCacheStatus[]
\end{mmaCell}
\begin{mmaCell}{Output}
False
\end{mmaCell}
}
{\tt \$cache} is a private boolean variable which determines if the output of certain calculations performed by {\tt amplituhedronBoundaries} is stored to file or not. The user can choose to have large calculations stored to file by setting the value of this variable to {\tt True} before loading the package: e.g.\
{\small
\begin{mmaCell}[index=1,moredefined={cache,amplituhedronBoundaries,printCacheStatus}]{Input}
\$cache=True;
<<amplituhedronBoundaries\`
printCacheStatus[]
\end{mmaCell}
\begin{mmaCell}[verbatimenv=]{Print}
	\includegraphics[width=5cm]{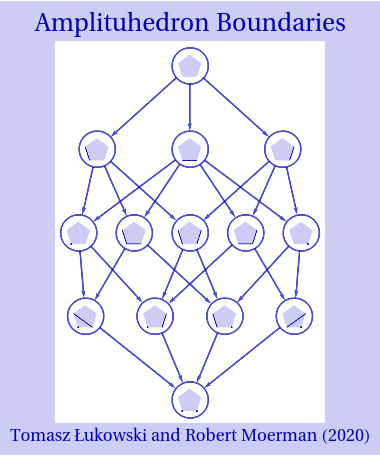}
\end{mmaCell}
\begin{mmaCell}{Output}
True
\end{mmaCell}
}

\item \mmaFuncDef{positiveMatrix}{{\mmaVarDef{nrows},\mmaVarDef{ncolumns}}} returns a generic $(\mmaVar{nrows}\times\mmaVar{ncolumns})$ matrix with all ordered maximal minors positive, provided $\mmaVar{nrows}\le\mmaVar{ncolumns}$.

\item \mmaFuncDef{orthComplement}{{\mmaVarDef{matrix}}} returns an orthogonal complement for the given \mmaVar{matrix}, provided that its number of rows is not greater than its number of columns.

\item \mmaFuncDef{topCell}{{\mmaVarDef{n},\mmaVarDef{k}}} returns the decorated permutation labelling the top cell of the positive Grassmannian $G_+(k,n)$.
{\small
\begin{mmaCell}[index=1,moredefined={topCell}]{Input}
topCell[4,2]
\end{mmaCell}
\begin{mmaCell}{Output}
\{3,4,5,6\}
\end{mmaCell}
}

\end{itemize}

\subsection{\emph{pos} -- positive Grassmannian boundaries}\label{sec:soft-pos}

\begin{itemize}
	\item \mmaFuncDef{posDimension}{\mmaVarDef{perm}} returns the dimension (in the positive Grassmannian) of the positroid cell labelled by the decorated permutation \mmaVar{perm} as computed\footnote{{\tt dimension} is a function in {\tt positroids} \cite{Bourjaily:2012gy}.} by {\tt dimension}.
	
	\item \mmaFuncDef{posBoundary}{\mmaVarDef{perm}} returns all co-dimension $1$ boundaries (in the positive Grassmannian) of the positroid cell labelled by the decorated permutation \mmaVar{perm} as computed\footnote{{\tt boundary} is a function in {\tt positroids} \cite{Bourjaily:2012gy}.} by {\tt boundary}. These boundaries are given by their decorated permutation labels.
	
	\item \mmaFuncDef{posBoundaries}{\mmaVarDef{codim},\mmaVarDef{perm}} returns all boundaries of co-dimension \mmaVar{codim} (in the positive Grassmannian) of the positroid cell labelled by the decorated permutation \mmaVar{perm}. These boundaries are given by their decorated permutation labels.
	
	\item \mmaFuncDef{posStratification}{\mmaVarDef{perm}} returns all boundaries in the positroid stratification of the cell labelled by the decorated permutation \mmaVar{perm}. These boundaries are given by their decorated permutation labels.
{\small
\begin{mmaCell}[index=1,moredefined={topCell,posStratification,Counts,posDimension,Length}]{Input}
posStratification[topCell[3,1]]
Counts[posDimension/@\%]
\end{mmaCell}
\begin{mmaCell}{Output}
{\{\{2,3,4\},\{1,3,5\},\{2,4,3\},\{3,2,4\},\{1,2,6\},\{1,5,3\},\newline\{4,2,3\}\}}
\end{mmaCell}
\begin{mmaCell}{Output}
<|2->1,1->3,0->3|>
\end{mmaCell}
The second line of the above output counts how many cells of a given dimension there are in this positroid stratification.
}

\item \mmaFuncDef{posInverseBoundary}{\mmaVarDef{perm}} returns all inverse boundaries (in the positive Grassmannian) of the positroid cell labelled by the decorated permutation \mmaVar{perm} as computed\footnote{{\tt inverseBoundary} is a function in {\tt positroids} \cite{Bourjaily:2012gy}.} by {\tt inverseBoundary}. 

\item \mmaFuncDef{posInverseBoundaries}{\mmaVarDef{dim},\mmaVarDef{perm}} returns all cells of dimension $\text{\tt posDimension[\mmaVar{perm}]+dim}$ (in the positive Grassmannian) which have as one of its boundaries the positroid cell labelled by the decorated permutation \mmaVar{perm}.
	
\item \mmaFuncDef{posInverseStratification}{\mmaVarDef{perm}} returns all inverse boundaries (in the positive Grassmannian) of the cell labelled by the decorated permutation \mmaVar{perm}.

\item \mmaFuncDef{posInterval}{{\mmaVarDef{perm1},\mmaVarDef{perm2}}} returns the interval of boundaries in the positive Grassmannian contained between the positroid cells labelled by the decorated permutations \mmaVar{perm1} and \mmaVar{perm2}. In particular, the function returns 
	\begin{itemize}
		\item all positroid cells contained in the intersection
		\begin{equation*}
		\text{\tt posInverseStratification[\mmaVar{perm1}]}\cap\,\text{\tt posStratification[\mmaVar{perm2}]}
		\end{equation*}
		if {\tt posDimension[\mmaVar{perm1}]}$<${\tt posDimension[\mmaVar{perm2}]},
		\item all positroid cells contained in the intersection \begin{equation*}
		\text{\tt posStratification[\mmaVar{perm1}]}\cap\,\text{\tt posInverseStratification[\mmaVar{perm2}]}
		\end{equation*}
		if {\tt posDimension[\mmaVar{perm1}]}$>${\tt posDimension[\mmaVar{perm2}]},
		\item {\tt\{\mmaVar{perm1}\}} if {\tt \mmaVar{perm1}}={\tt\mmaVar{perm2}},
		\item an empty list {\tt\{\}} otherwise.
	\end{itemize}
{\small
\begin{mmaCell}[index=1,moredefined={posInterval}]{Input}
posInterval[\{2,3,4\},\{1,2,6\}]
\end{mmaCell}
\begin{mmaCell}{Output}
\{\{1,2,6\},\{1,3,5\},\{2,3,4\},\{3,2,4\}\}
\end{mmaCell}
}

\item \mmaFuncDef{posPoset}{\mmaVarDef{perm}} returns the (transitively reduced) adjacency matrix for the poset\footnote{The (strict) partial order $\prec_C$ is induced via inclusion with respect to the boundary operator: $\pi'\prec_C\pi \iff \pi'\in\partial_C\pi$.} of all boundaries in the positive Grassmannian of the positroid cell labelled by the decorated permutation \mmaVar{perm}.
{\small
\begin{mmaCell}[index=1,moredefined={topCell,posPoset,MatrixForm,AdjacencyGraph}]{Input}
posPoset[topCell[3,1]];
\{MatrixForm[\%],AdjacencyGraph[\%]\}
\end{mmaCell}
%
\begin{mmaCell}[verbatimenv=]{Output}
	\raisebox{-0.5\totalheight}{\includegraphics[width=8cm]{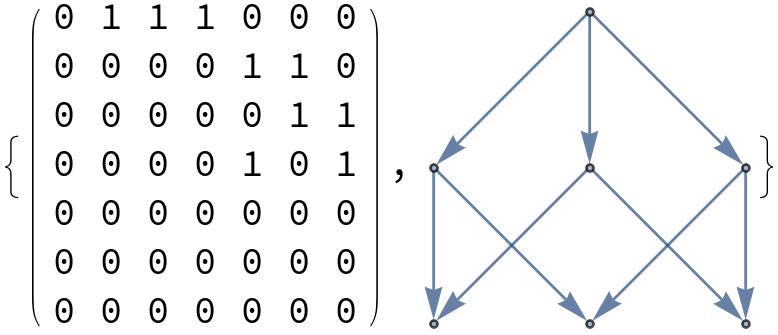}}
\end{mmaCell}
}

\item \mmaFuncDef{posPoset}{{\mmaVarDef{perm1},\mmaVarDef{perm2}}} returns the (transitively reduced) adjacency matrix which for the poset of all boundaries in the interval {\tt posInterval[\mmaVar{perm1},\mmaVar{perm2}]}.
{\small
\begin{mmaCell}[index=1,moredefined={posPoset,MatrixForm,AdjacencyGraph}]{Input}
posPoset[\{2,3,4\},\{1,2,6\}];
\{MatrixForm[\%],AdjacencyGraph[\%]\}	
\end{mmaCell}
\begin{mmaCell}[verbatimenv=]{Output}
	\raisebox{-0.5\totalheight}{\includegraphics[width=4cm]{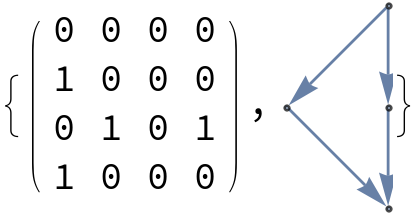}}
\end{mmaCell}
}

\item \mmaFuncDef{posPermToLeDiagram}{\mmaVarDef{perm}} returns the \Le-diagram \cite{Postnikov:2006kva} corresponding to the decorated permutation \mmaVar{perm}.
{\small
\begin{mmaCell}[index=1,moredefined={posPermToLeDiagram,posStratification,topCell}]{Input}
posPermToLeDiagram/@posStratification[topCell[3,1]]
\end{mmaCell}
\begin{mmaCell}[verbatimenv=]{Output}
\raisebox{-0.5\totalheight}{\includegraphics[width=\linewidth]{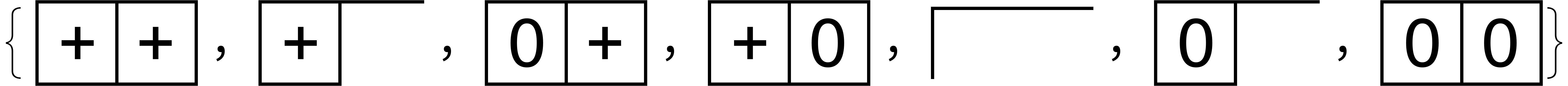}}
\end{mmaCell}
}

\item \mmaFuncDef{posLeDiagramToPerm}{\mmaVarDef{lediagram}} returns the decorated permutation corresponding to the \Le-diagram \cite{Postnikov:2006kva} \mmaVar{lediagram}.
{\small
	\begin{mmaCell}[index=1,verbatimenv=]{Input}
\raisebox{-0.5\totalheight}{\includegraphics[width=\linewidth]{posPermToLeDiagram-3-1.png}}\texttt{;}\\[0.5em]\texttt{posLeDiagramToPerm/@\%}
	\end{mmaCell}
	\begin{mmaCell}{Output}
\{\{2,3,4\},\{1,3,5\},\{2,4,3\},\{3,2,4\},\{1,2,6\},\{1,5,3\},
	\{4,2,3\}\}
	\end{mmaCell}
}
\end{itemize}

\subsection{\emph{amp} -- amplituhedron boundaries}\label{sec:soft-amp}

\begin{itemize}

\item \mmaFuncDef{ampDimension}{\mmaVarDef{m},\mmaVarDef{perm}} returns the dimension of the image\footnote{We wish to thank Jacob Bourjaily for useful discussions about how to implement this function efficiently.} (in the amplituhedron $\mathcal{A}^{(\mmaVar{m})}$) of the positroid cell labelled by the decorated permutation \mmaVar{perm}. 
	
If the amplituhedron dimension for this positroid cell (and in fact all cells in the positroid stratification of the same positive Grassmannian) is stored to file in the {\tt Data/} folder, then the amplituhedron dimension will be retrieved directly from the file and thereafter stored in the kernel's memory. The first time the amplituhedron dimension is retrieved from the file, the following message will be displayed:
{\small
\begin{mmaCell}[index=1,moredefined={topCell,ampDimension}]{Input}
topCell[4,1]//ampDimension[2]
\end{mmaCell}
\begin{mmaCell}{Message}
ampDimension: called for the first time with n=4, k=1, m=2; 
	loading definition from ``<notebook directory>/Data/
	ampDimension-4-1-2.m''.
\end{mmaCell}
\begin{mmaCell}{Output}
2
\end{mmaCell}
}

If the amplituhedron dimension for this positroid cell is \emph{not} stored to file in the {\tt Data/} folder, the behaviour of \mmaFuncName{ampDimension} depends on the value of {\tt \$cache}.

\begin{itemize}
	\item {\tt \$cache=False}: the amplituhedron dimension of the poistroid cell is calculated directly and thereafter stored in the kernel's memory.
	{\small
		\begin{mmaCell}[index=1,moredefined={topCell,ampDimension}]{Input}
topCell[4,1]//ampDimension[2]
		\end{mmaCell}
		\begin{mmaCell}{Message}
ampDimension: called for the first time with n=4, k=1, m=2; 
	calculating definition on-the-fly.
		\end{mmaCell}
		\begin{mmaCell}{Output}
2
		\end{mmaCell}
	}
	
	\item {\tt \$cache=True}: the amplituhedron dimension of \emph{all} cells in the positroid stratification of the same positive Grassmannian is first calculated, stored in the kernel's memory and stored to file in the {\tt Data/} folder before returning the amplituhedron dimension of the positroid cell of interest.\\
	{\small
		\begin{mmaCell}[index=1,moredefined={topCell,ampDimension}]{Input}
topCell[4,1]//ampDimension[2]
		\end{mmaCell}
		\begin{mmaCell}{Message}
ampDimension: called for the first time with n=4, k=1, m=2; 
	constructing and saving definition to 
	``<notebook directory>/Data/ampDimension-4-1-2.m''.
		\end{mmaCell}
		\begin{mmaCell}{Output}
2
		\end{mmaCell}
	}
\end{itemize}

\mmaFunc{ampDimension}{\mmaVar{m},\mmaVar{perm}} can be called using \\\mmaFunc{ampDimension}{{\mmaVar{perm},m$\to$\mmaVar{m}}} or simply \mmaFunc{ampDimension}{\mmaVar{perm}} where the default is {\tt m$\to$2}. 

\item \mmaFuncDef{ampFaceQ}{\mmaVarDef{m},\mmaVarDef{perm}} returns {\tt True} if \mmaVar{perm} labels an amplituhedron face, and otherwise {\tt False}. We say that $\pi=\mmaVar{perm}$ labels an amplituhedron face if there are no cells in the inverse boundary of \mmaVar{perm} with the same amplituhedron dimension as \mmaVar{perm}:
	\begin{align*}
	\forall_{ \pi'\in\partial^{-1}_C\pi}:\dim_A(\pi')>\dim_A(\pi).
	\end{align*}
	
\mmaFunc{ampFaceQ}{\mmaVar{m},\mmaVar{perm}} can be called using \\\mmaFunc{ampFaceQ}{{\mmaVar{perm},m$\to$\mmaVar{m}}} or simply \mmaFunc{ampFaceQ}{\mmaVar{perm}} where the default is {\tt m$\to$2}. 

\item \mmaFuncDef{ampFaces}{\mmaVarDef{m},\mmaVarDef{perm}} returns all cells in {\tt posStratification[\mmaVar{perm}]} which are amplituhedron faces.
	
\mmaFunc{ampFaces}{\mmaVar{m},\mmaVar{perm}} can be called using \mmaFunc{ampFaces}{\mmaVar{perm},{m$\to$\mmaVar{m}}} or simply \mmaFunc{ampFaces}{\mmaVar{perm}} where the default is {\tt m$\to$2}. 

\item \mmaFuncDef{ampBoundaries}{{\mmaVarDef{m},\mmaVarDef{codim}},\mmaVarDef{perm}} returns all boundaries of co-dimension \mmaVar{codim} in the amplituhedron $\mathcal{A}^{(\mmaVar{m})}$ of the cell labelled by the decorated permutation \mmaVar{perm} provided it is either a top cell or a boundary of the top cell in the amplituhedron.
	
\mmaFunc{ampBoundaries}{{\mmaVar{m},\mmaVar{codim}}\mmaVar{perm}} can be called using \\\mmaFunc{ampBoundaries}{{\mmaVar{codim},m$\to$\mmaVar{m}},\mmaVar{perm}} or simply \\\mmaFunc{ampBoundaries}{\mmaVar{codim},\mmaVar{perm}} where the default is {\tt m$\to$2}. 

\item \mmaFuncDef{ampStratification}{\mmaVarDef{m},\mmaVarDef{perm}} returns the stratification of all boundaries of the cell labelled by the decorated permutation \mmaVar{perm} in the amplituhedron $\mathcal{A}^{(\mmaVar{m})}$ provided \mmaVar{perm} labels either a top cell or a boundary of the top cell in the amplituhedron.
	
	The first-time behaviour of this function depends on whether or not its output already exists on file in the {\tt Data/} folder. If this is the case, then this output is simply retrieved and stored in the kernel's memory. Otherwise the output is calculated directly and stored in the kernel's memory, and if {\tt \$cache=True} then the output is also then stored to file. 
	
	A message is displayed the first time the function is executed analogous to the messages displayed by {\tt ampDimension} detailing its first-time behaviour. 
		
	\mmaFunc{ampStratification}{\mmaVar{m},\mmaVar{perm}} can be called using \\\mmaFunc{ampStratification}{{\mmaVar{perm},m$\to$\mmaVar{m}}} or simply \\\mmaFunc{ampStratification}{\mmaVar{perm}} where the default is {\tt m$\to$2}.
	
	The following list of functions use \mmaFuncName{ampStratification} in its computation:
	\begin{itemize}
		\item {\tt ampInverseStratification}
		\item {\tt ampBoundary}
		\item {\tt ampInverseBoundary}
		\item {\tt ampInterval}
		\item {\tt ampPoset}
		\item {\tt ampPermToGraphic}
		\item {\tt ampStratificationToHasse}
		\item {\tt ampIntervalToHasse}
		\item {\tt ampStratificationTo3D}	
		\item {\tt ampFacetsToGraph}
	\end{itemize}
	So when calling any of these functions for the first time, one of the above messages generated by {\tt ampStratification} may be displayed.
	
	\item \mmaFuncDef{ampInverseStratification}{\mmaVarDef{m},\mmaVarDef{perm}} returns the stratification of all inverse boundaries of the cell labelled by the decorated permutation \mmaVar{perm} in the amplituhedron $\mathcal{A}^{(\mmaVar{m})}$ provided \mmaVar{perm} labels a boundary in the amplituhedron. 
		
	\mmaFunc{ampInverseStratification}{\mmaVar{m}\mmaVar{perm}} can be called using \\\mmaFunc{ampInverseStratification}{{\mmaVar{perm},m$\to$\mmaVar{m}}} or simply \\\mmaFunc{ampInverseStratification}{\mmaVar{perm}} where the default is {\tt m$\to$2}.
	
	\item \mmaFuncDef{ampBoundary}{\mmaVarDef{m},\mmaVarDef{perm}} returns all co-dimension 1 boundaries (in the amplituhedron $\mathcal{A}^{(\mmaVar{m})}$) of the cell labelled by the decorated permutation \mmaVar{perm} provided \mmaVar{perm} labels a boundary in the amplituhedron.
		
	\emph{Currently \mmaFuncName{ampBoundary} is only implemented for {\tt m=2}.}
		
	\mmaFunc{ampBoundary}{2,\mmaVar{perm}} can be called using \mmaFunc{ampBoundary}{{\mmaVar{perm},m$\to$2}} or simply \mmaFunc{ampBoundary}{\mmaVar{perm}} where the default is {\tt m$\to$2}.
	
\item \mmaFuncDef{ampInverseBoundary}{\mmaVarDef{m},\mmaVarDef{perm}} returns all inverse boundaries (in the amplituhedron with $\mathcal{A}^{(\mmaVar{m})}$) with amplituhedron dimension \\$\text{\tt ampDimension[\mmaVar{m}][\mmaVar{perm}]+1}$	of the cell labelled by the decorated permutation \mmaVar{perm} provided \mmaVar{perm} labels a boundary in the amplituhedron.
		
		\emph{Currently \mmaFuncName{ampInverseBoundary} is only implemented for {\tt m=2}.}
		
		\mmaFunc{ampInverseBoundary}{2,\mmaVar{perm}} can be called using \\\mmaFunc{ampInverseBoundary}{{\mmaVar{perm},m$\to$2}} or simply \\\mmaFunc{ampInverseBoundary}{\mmaVar{perm}} where the default is {\tt m$\to$2}.
		
\item \mmaFuncDef{ampInterval}{\mmaVarDef{m},{\mmaVarDef{perm1},\mmaVarDef{perm2}}} returns the interval of boundaries in the amplituhedron $\mathcal{A}^{(m)}$ contained between the cells labelled by the decorated permutations \mmaVar{perm1} and \mmaVar{perm2}.
	
	\emph{Currently \mmaFuncName{ampInterval} is only implemented for {\tt m=2}.}
	
	\mmaFunc{ampInterval}{2,\mmaVar{perm1},\mmaVar{perm2}} can be called using \\\mmaFunc{ampInterval}{{\mmaVar{perm1},\mmaVar{perm2},{\tt m$\to$2}}} or simply \\\mmaFunc{ampInterval}{{\mmaVar{perm1},\mmaVar{perm2}}} where the default is {\tt m$\to$2}.
	
\item	\mmaFuncDef{ampPoset}{\mmaVarDef{m},\mmaVarDef{perm}} returns the (transitively reduced) adjacency matrix which defines the poset structure induced by inclusion for all boundaries (in the amplituhedron $\mathcal{A}^{(m)}$) of the cell labelled by the decorated permutation \mmaVar{perm}.
		
		\emph{Currently \mmaFuncName{ampPoset} is only implemented for {\tt m=2}.}
		
		\mmaFunc{ampPoset}{2,\mmaVar{perm}} can be called using \mmaFunc{ampPoset}{{\mmaVar{perm},{\tt m$\to$2}}} or simply \mmaFunc{ampPoset}{\mmaVar{perm}} where the default is {\tt m$\to$2}.
		
\item \mmaFuncDef{ampPoset}{\mmaVarDef{m},{\mmaVarDef{perm1},\mmaVarDef{perm2}}} returns the (transitively reduced) adjacency matrix which defines the poset structure induced by inclusion for all boundaries in the interval {\tt ampInterval[\mmaVar{m}][\mmaVar{perm1},\mmaVar{perm2}]}.
			
			\emph{Currently \mmaFuncName{ampPoset} is only implemented for {\tt m=2}.}
			
			\mmaFunc{ampPoset}{2,{\mmaVar{perm1},\mmaVar{perm2}}} can be called using \mmaFunc{ampPoset}{{\mmaVar{perm1},\\\mmaVar{perm2},{\tt m$\to$2}}} or simply \mmaFunc{ampPoset}{{\mmaVar{perm1},\mmaVar{perm2}}} where the default is {\tt m$\to$2}.
			
\item \mmaFuncDef{ampGeneralizedTriangles}{\mmaVarDef{m},{\mmaVarDef{n},\mmaVarDef{k}}} returns all generalized triangles in the amplituhedron $\mathcal{A}^{(\mmaVar{m})}_{\mmaVar{n},\mmaVar{k}}$ \cite{Lukowski:2019sxw} -- all decorated permutation labels $\pi$ for positroid cells $C_\pi\in G_+(\mmaVar{k},\mmaVar{n})$ satisfying 
	\begin{equation*}
	\dim_C(\pi)=\dim_A(\pi)=\mmaVar{m}\cdot\mmaVar{k},
	\end{equation*}
	and\footnote{{\tt intersectionNumber} is a function in {\tt positroids} \cite{Bourjaily:2012gy}.} {\tt intersectionNumber[\mmaVar{perm},\mmaVar{m}]==1}.

	The first-time behaviour of this function depends on whether or not its output already exists on file in the {\tt Data/} folder. If this is the case, then this output is simply retrieved and stored in the kernel's memory. Otherwise the output is calculated directly and stored in the kernel's memory, and if {\tt \$cache=True} then the output is also then stored to file. 
	
	A message is displayed the first time the function is executed analogous to the messages displayed by {\tt ampDimension} detailing its first-time behaviour.

	\mmaFunc{ampGeneralizedTriangles}{\mmaVar{m},{\mmaVar{n},\mmaVar{k}}} can be called using \\\mmaFunc{ampGeneralizedTriangles}{{\mmaVar{n},\mmaVar{k},{\tt m$\to$\mmaVar{m}}}} or simply \\\mmaFunc{ampGeneralizedTriangles}{{\mmaVar{n},\mmaVar{k}}} where the default is {\tt m$\to$2}.
	
\item \mmaFuncDef{ampPermToGraphic}{\mmaVarDef{m},\mmaVarDef{perm}} returns the compact graphical label for the cell in the amplituhedron $\mathcal{A}^{(\mmaVar{m})}$ corresponding to the decorated permutation \mmaVar{perm}.
	
	\emph{Currently \mmaFuncName{ampPermToGraphic} is only implemented for {\tt m=2}.} These compact graphical labels for {\tt m=2} are defined in \cite{Lukowski:2019kqi}: they consist of lines and points inside an $n$-gon (where $n$ is given by the length of \mmaVar{perm}).
	{\small
		\begin{mmaCell}[index=1,moredefined={ampPermToGraphic,ampStratification,topCell}]{Input}
ampPermToGraphic[2]/@ampStratification[2][topCell[3,1]]
		\end{mmaCell}
		\begin{mmaCell}[verbatimenv=]{Output}
\raisebox{-0.5\totalheight}{\includegraphics[width=\linewidth]{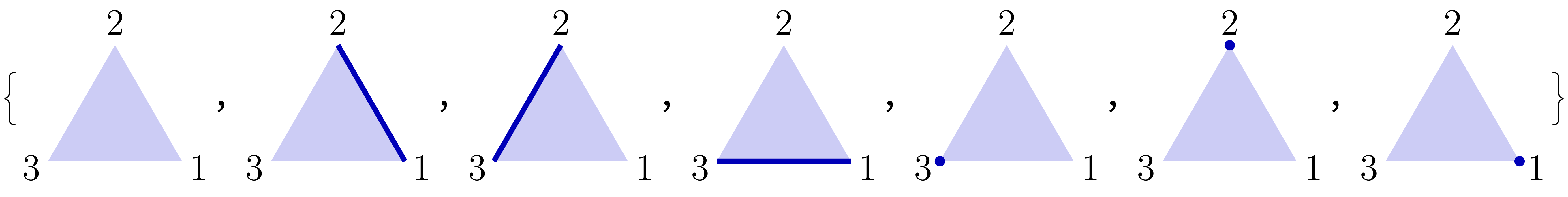}}
		\end{mmaCell}
	}
	
	\mmaFunc{ampPermToGraphic}{2,\mmaVar{perm}} can be called using \\\mmaFunc{ampPermToGraphic}{{\mmaVar{perm},{\tt m$\to$2}}} or simply \\\mmaFunc{ampPermToGraphic}{\mmaVar{perm}} where the default is {\tt m$\to$2}.
	
	\mmaFuncName{ampPermToGraphic} additionally admits the following optional parameters:\\\\
	\mbox{\hspace{-1.5cm}\begin{tabular}{P{4cm}@{$\,$}c@{$\,$}P{2cm}@{$~$}c@{$~$}P{8.5cm}}
			\hline
			{Option}&&{Value}&&{Description}\\\hline\hline 
			{\tt showLabels}&$\ast$&{\tt True}&{$\triangleright$}&displays the labels of the vertices of the $n$-gon\\&&{\tt False}&{$\triangleright$}&leaves the labels of the vertices of the $n$-gon off\\\hline
			{\tt noLabelsResize}&$\ast$&{\tt True}&{$\triangleright$}&resizes the $n$-gon if labels are not shown\\&&{\tt False}&{$\triangleright$}&the $n$-gon is not resized if labels are not shown\\\hline
			{\tt imageSize}&$\ast$&{\tt 80}&{$\triangleright$}&the image size can be set arbitrarily, but defaults to {\tt 80}\\\hline
	\end{tabular}}
	
	For each optional parameter, we have indicated its default value by an asterix ``$\ast$''.
	{\small
		\begin{mmaCell}[index=1,moredefined={ampPermToGraphic,ampStratification,topCell}]{Input}
topCell[3,1]//ampPermToGraphic[2,showLabels->False];
topCell[3,1]//ampPermToGraphic[2,showLabels->False, 
	noLabelsResize->True];
topCell[3,1]//ampPermToGraphic[2,imageSize->50];
\{\%\%\%,\%\%,\%\}
		\end{mmaCell}
		\begin{mmaCell}[verbatimenv=]{Output}
\raisebox{-0.5\totalheight}{\includegraphics{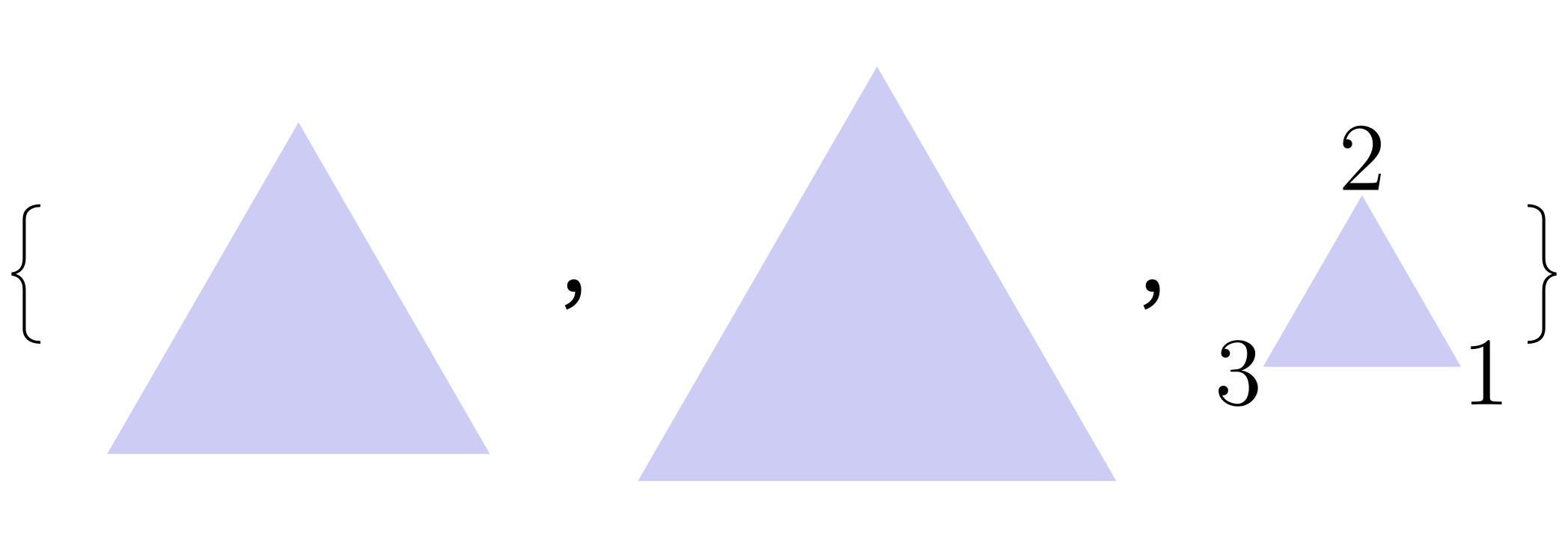}}
		\end{mmaCell}
	}

\item \mmaFuncDef{ampStratificationToHasse}{\mmaVarDef{m},\mmaVarDef{perm}} returns a Hasse diagram for the poset of boundaries in {\tt ampStratification[\mmaVar{m}][\mmaVar{perm}]}.
	
	\emph{Currently \mmaFuncName{ampStratificationToHasse} is only implemented for {\tt m=2}.}
{\small
	\begin{mmaCell}[index=1,moredefined={ampStratificationToHasse,topCell}]{Input}
ampStratificationToHasse[2][topCell[4,1]]
	\end{mmaCell}
	\begin{mmaCell}[verbatimenv=]{Output}
\raisebox{-0.5\totalheight}{\includegraphics[width=8cm]{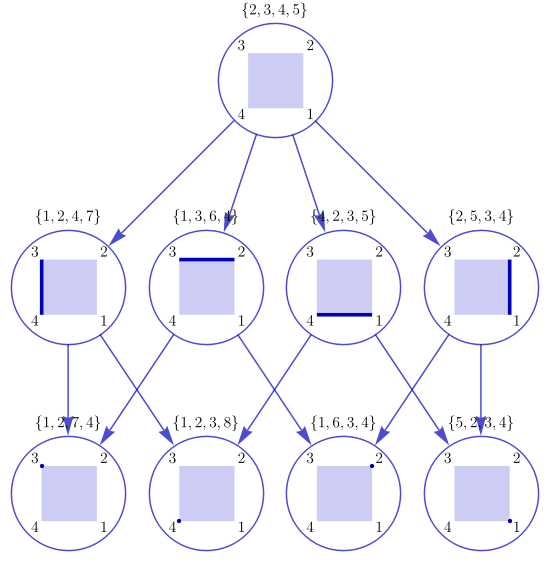}}
	\end{mmaCell}
}
	
	\mmaFunc{ampStratificationToHasse}{2,\mmaVar{perm}} can be called using \\\mmaFunc{ampStratificationToHasse}{{\mmaVar{perm},{\tt m$\to$2}}} or simply \\\mmaFunc{ampStratificationToHasse}{\mmaVar{perm}} where the default is {\tt m$\to$2}.
	
	\mmaFuncName{ampStratificationToHasse} additionally admits the following optional parameters:\\\\
	\mbox{\hspace{-1.5cm}\begin{tabular}{P{4cm}@{$\,$}c@{$\,$}P{2cm}@{$~$}c@{$~$}P{8.5cm}}
			\hline
			{Option}&&{Value}&&{Description}\\\hline\hline 
			{\tt showGraphics}&$\ast$&{\tt True}&{$\triangleright$}&displays graphical labels generated by {\tt ampPermToGraphic} in the nodes of the Hasse diagram\\&&{\tt False}&{$\triangleright$}&leaves the nodes of the Hasse diagram empty\\\hline
			{\tt showLabels}&$\ast$&{\tt True}&{$\triangleright$}&displays the labels of the vertices of the $n$-gon in the graphic generated by {\tt ampPermToGraphic}\\&&{\tt False}&{$\triangleright$}&leaves the labels of the vertices of the $n$-gon in the graphic generated by {\tt ampPermToGraphic} off\\\hline
			{\tt showPermutations}&$\ast$&{\tt True}&{$\triangleright$}&displays permutation labels above the nodes of the Hasse diagram\\&&{\tt False}&{$\triangleright$}&leaves permutation labels above the nodes of the Hasse diagram off\\\hline
			{\tt rasterize}&$\ast$&{\tt True}&{$\triangleright$}&rasterizes the Hasse diagram so that is can be easily resized\\&&{\tt False}&{$\triangleright$}&does not rasterize the Hasse diagram\\\hline
			{\tt imageResolution}&$\ast$&{\tt Automatic}&{$\triangleright$}&the resolution of the rasterized Hasse diagram can be set arbitrarily, but defaults to {\tt Automatic}\\\hline
			{\tt edgeScale}&$\ast$&{\tt 1}&{$\triangleright$}&the edge thickness can be scaled arbitrarily, but defaults to {\tt 1}\\\hline
	\end{tabular}}
	
	For each optional parameter, we have indicated its default value by an asterix ``$\ast$''.
	
\item \mmaFuncDef{ampStratificationToTable}{\mmaVarDef{m},\mmaVarDef{perm}} returns a table of the cells in {\tt ampStratification[\mmaVar{m}][\mmaVar{perm}]} organised with respect to their amplituhedron dimensions.
	
	\emph{Currently \mmaFuncName{ampStratificationToTable} is only implemented for {\tt m=2}.}
{\small
	\begin{mmaCell}[index=1,moredefined={ampStratificationToTable,topCell}]{Input}
topCell[5,1]//ampStratificationToTable[2,
	fancyPermutations->True]
	\end{mmaCell}
	\begin{mmaCell}[verbatimenv=]{Output}
		\raisebox{-0.5\totalheight}{\includegraphics[width=0.9\linewidth]{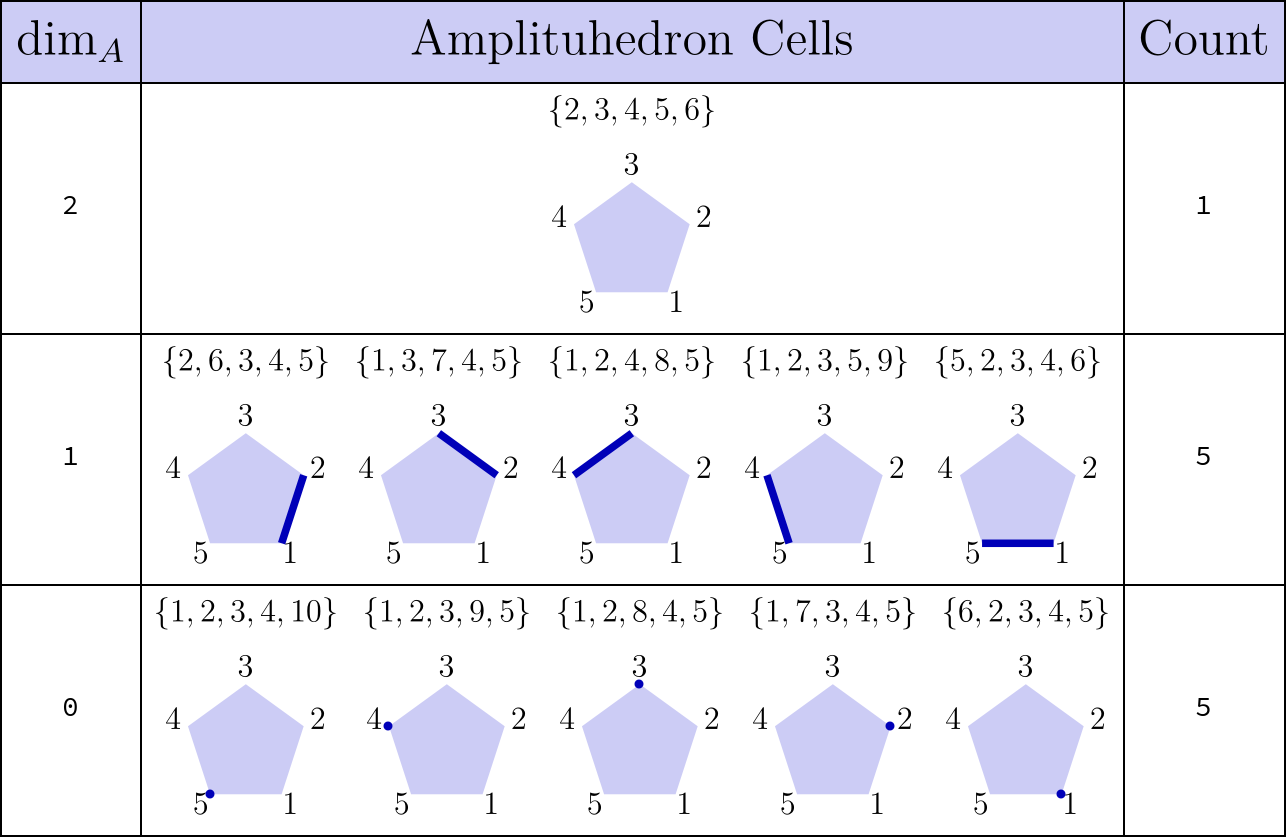}}
	\end{mmaCell}
}
	
	\mmaFunc{ampStratificationToTable}{2,\mmaVar{perm}} can be called using \\\mmaFunc{ampStratificationToTable}{{\mmaVar{perm},m$\to$2}} or simply \\\mmaFunc{ampStratificationToTable}{\mmaVar{perm}} where the default is {\tt m$\to$2}.
	
	\mmaFuncName{ampStratificationToTable} additionally admits the following optional parameters:\\\\
	\mbox{\hspace{-1.5cm}\begin{tabular}{P{4cm}@{$\,$}c@{$\,$}P{2.4cm}@{$~$}c@{$~$}P{8.1cm}}
			\hline
			{Option}&&{Value}&&{Description}\\\hline\hline 
			{\tt showLabels}&$\ast$&{\tt True}&{$\triangleright$}&displays the labels of the vertices of the $n$-gon in the graphic generated by {\tt ampPermToGraphic}\\&&{\tt False}&{$\triangleright$}&leaves the labels of the vertices of the $n$-gon in the graphic generated by {\tt ampPermToGraphic} off\\\hline
			{\tt showPermutations}&$\ast$&{\tt True}&{$\triangleright$}&displays permutation labels above the graphics displayed in the table\\&&{\tt False}&{$\triangleright$}&leaves permutation labels above the graphics displayed in the table off\\\hline
			{\tt fancyPermutations}&&{\tt True}&{$\triangleright$}&displays permutation labels in \LaTeX~font.\\&$\ast$&{\tt False}&{$\triangleright$}&leaves permutation labels unformatted.\\\hline
			{\tt graphicType}&$\ast$&{\tt "polygon"}&{$\triangleright$}&displays the graphic generated by {\tt ampPermToGraphic} for each boundary.\\&&{\tt "lediagram"}&{$\triangleright$}&displays the \Le-diagram generated by {\tt posPermToLeDiagram} for each boundary.\\\hline
	\end{tabular}}

	For each optional parameter, we have indicated its default value by an asterix ``$\ast$''.
	
\item \mmaFuncDef{ampIntervalToHasse}{\mmaVarDef{m},{\mmaVarDef{perm1},\mmaVarDef{perm2}}} returns a Hasse diagram for the poset of boundaries in {\tt ampInterval[\mmaVar{m}][\mmaVar{perm1},\mmaVar{perm2}]}.
		
		\emph{Currently \mmaFuncName{ampIntervalToHasse} is only implemented for {\tt m=2}.}
		
		\mmaFunc{ampIntervalToHasse}{2,{\mmaVar{perm1},\mmaVar{perm2}}} can be called using \\\mmaFunc{ampIntervalToHasse}{{\mmaVar{perm1},\mmaVar{perm2},{\tt m$\to$2}}} or simply \\\mmaFunc{ampIntervalToHasse}{{\mmaVar{perm1},\mmaVar{perm2}}} where the default is {\tt m$\to$2}.
		
		\mmaFuncName{ampIntervalToHasse} additionally admits the same optional parameters as {\tt ampStratificationToHasse}.
		
\item \mmaFuncDef{ampStratificationTo3D}{\mmaVarDef{m}\mmaVarDef{perm}} returns a graph of connections between zero-dimensional cells in \\{\tt ampStratification[\mmaVar{m}][\mmaVar{perm}]} provided the cell in the amplituhedron $\mathcal{A}^{(\mmaVar{m})}$ labelled by the decorated permutation \mmaVar{perm} has amplituhedron dimension less than or equal to $3$.
	
	\emph{Currently \mmaFuncName{ampStratificationTo3D} is only implemented for {\tt m=2}.}	
	{\small
		\begin{mmaCell}[index=1,moredefined={ampStratificationTo3D}]{Input}
\{2,4,5,7\}//ampStratificationTo3D[2]
		\end{mmaCell}
		\begin{mmaCell}[verbatimenv=]{Output}
			\raisebox{-0.5\totalheight}{\includegraphics[width=5cm]{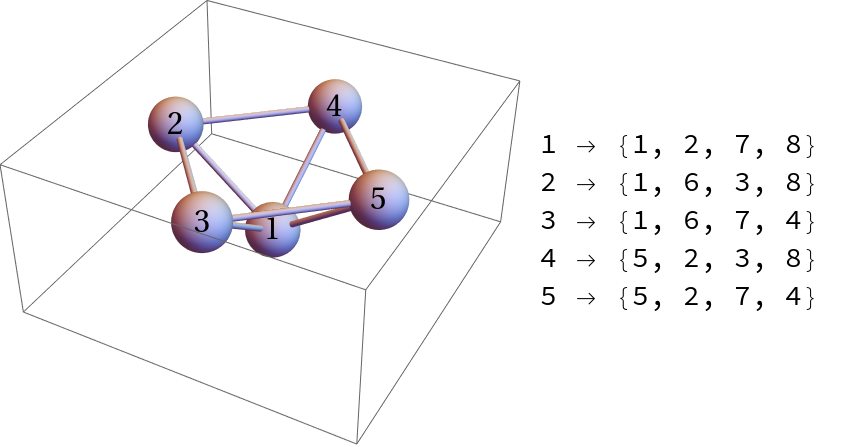}}
		\end{mmaCell}
	}
	
	\mmaFunc{ampStratificationTo3D}{2,\mmaVar{perm}} can be called using \\\mmaFunc{ampStratificationTo3D}{{\mmaVar{perm},m$\to$2}} or simply \\\mmaFuncName{ampStratificationTo3D}{\mmaVar{perm}} where the default is {\tt m$\to$2}.
	
	\mmaFuncName{ampStratificationTo3D} additionally admits the following optional parameters:\\\\
	\mbox{\hspace{-1.5cm}\begin{tabular}{P{4cm}@{$\,$}c@{$\,$}P{2cm}@{$~$}c@{$~$}P{8.5cm}}
			\hline
			{Option}&&{Value}&&{Description}\\\hline\hline 
			{\tt showGraphics}&&{\tt True}&{$\triangleright$}&displays graphical labels generated by {\tt ampPermToGraphic} for each vertex in the graph\\&$\ast$&{\tt False}&{$\triangleright$}&does not display graphical labels\\\hline
			{\tt showLabels}&&{\tt True}&{$\triangleright$}&displays the labels of the vertices of the $n$-gon in the graphic generated by {\tt ampPermToGraphic}\\&$\ast$&{\tt False}&{$\triangleright$}&leaves the labels of the vertices of the $n$-gon in the graphic generated by {\tt ampPermToGraphic} off\\\hline
			{\tt showPermutations}&$\ast$&{\tt True}&{$\triangleright$}&displays permutation labels for each vertex in the graph\\&&{\tt False}&{$\triangleright$}&does not display permutation labels\\\hline
			{\tt imageSize}&$\ast$&{\tt 40}&{$\triangleright$}&the image size of the graphic generated by {\tt ampPermToGraphic} can be set arbitrarily, but defaults to {\tt 40}\\\hline
	\end{tabular}}
	
	For each optional parameter, we have indicated its default value by an asterix ``$\ast$''.

\item \mmaFuncDef{ampFacetsToGraph}{\mmaVarDef{m}\mmaVarDef{perm}} returns a star-shaped graph of the facets (co-dimension $1$ boundaries) of the cell in the amplituhedron $\mathcal{A}^{(\mmaVar{m})}$ labelled by the decorated permutation \mmaVar{perm} provided it is an element of the boundary stratification of the amplituhedron and it has amplituhedron dimension greater than or equal to 1.
	
	\emph{Currently \mmaFuncName{ampFacetsToGraph} is only implemented for {\tt m=2}.}
	{\small
		\begin{mmaCell}[index=1,moredefined={ampFacetsToGraph,topCell}]{Input}
topCell[6,2]//ampFacetsToGraph[2]
		\end{mmaCell}
		\begin{mmaCell}[verbatimenv=]{Output}
			\raisebox{-0.5\totalheight}{\includegraphics[width=5cm]{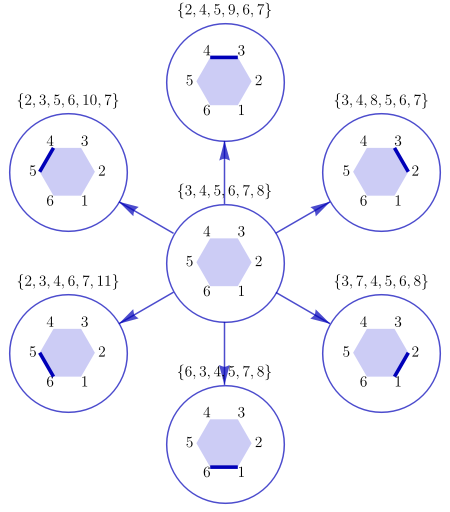}}
		\end{mmaCell}
	}
	
	\mmaFunc{ampFacetsToGraph}{2,\mmaVar{perm}} can be called using \\\mmaFunc{ampFacetsToGraph}{{\mmaVar{perm},m$\to$2}} or simply \\\mmaFunc{ampFacetsToGraph}{\mmaVar{perm}} where the default is {\tt m$\to$2}.
	
	\mmaFuncName{ampFacetsToGraph} additionally admits the same optional parameters as {\tt ampStratificationToHasse}.
\end{itemize}

\subsection{\emph{mom} -- momentum amplituhedron boundaries}\label{sec:soft-mom}
All functions below with the prefix \emph{mom} can only be called with {\tt m$\,\in\,$\{2,4\}}; the default value is {\tt m$\to$2}.

\begin{itemize}
%

\item \mmaFuncDef{momStratificationToHasse}{\mmaVarDef{m}\mmaVarDef{perm}} returns a Hasse diagram for the poset of boundaries in {\tt momStratification[\mmaVar{m}][\mmaVar{perm}]}.
{\small
	\begin{mmaCell}[index=1,moredefined={momStratificationToHasse}]{Input}
\{2,3,4\}//momStratificationToHasse[2]
	\end{mmaCell}
	\begin{mmaCell}[verbatimenv=]{Output}
\raisebox{-0.5\totalheight}{\includegraphics[width=4cm]{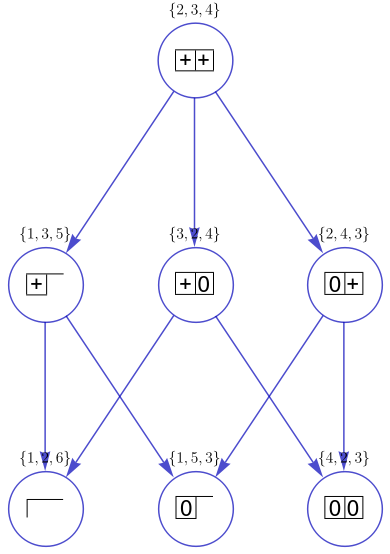}}
	\end{mmaCell}
}
	
	\mmaFunc{momStratificationToHasse}{\mmaVar{m},\mmaVar{perm}} can be called using \\\mmaFunc{momStratificationToHasse}{{\mmaVar{perm},m$\to$\mmaVar{m}}} or simply \\\mmaFunc{momStratificationToHasse}{\mmaVar{perm}} where the default is {\tt m$\to$2}.
	
	\mmaFuncName{momStratificationToHasse} additionally admits the following optional parameters:\\\\
	\mbox{\hspace{-1.5cm}\begin{tabular}{P{4cm}@{$\,$}c@{$\,$}P{2cm}@{$~$}c@{$~$}P{8.5cm}}
			\hline
			{Option}&&{Value}&&{Description}\\\hline\hline 
			{\tt showGraphics}&$\ast$&{\tt True}&{$\triangleright$}&displays \Le-diagrams in the nodes of the Hasse diagram\\&&{\tt False}&{$\triangleright$}&leaves the nodes of the Hasse diagram empty\\\hline
			{\tt showPermutations}&$\ast$&{\tt True}&{$\triangleright$}&displays permutation labels above the nodes of the Hasse diagram\\&&{\tt False}&{$\triangleright$}&leaves permutation labels above the nodes of the Hasse diagram off\\\hline
			{\tt rasterize}&$\ast$&{\tt True}&{$\triangleright$}&rasterizes the Hasse diagram so that is can be easily resized\\&&{\tt False}&{$\triangleright$}&does not rasterize the Hasse diagram\\\hline
			{\tt imageResolution}&$\ast$&{\tt Automatic}&{$\triangleright$}&the resolution of the rasterized Hasse diagram can be set arbitrarily, but defaults to {\tt Automatic}\\\hline
			{\tt edgeScale}&$\ast$&{\tt 1}&{$\triangleright$}&the edge thickness can be scaled arbitrarily, but defaults to {\tt 1}\\\hline
	\end{tabular}}
	
	For each optional parameter, we have indicated its default value by an asterix ``$\ast$''.
	
\item \mmaFuncDef{momStratificationToTable}{\mmaVarDef{m},\mmaVarDef{perm}} returns a table of the cells in {\tt momStratification[\mmaVar{m}][\mmaVar{perm}]} organised with respect to their momentum amplituhedron dimensions.
{\small
	\begin{mmaCell}[index=1,moredefined={topCell,momStratificationToTable}]{Input}
topCell[5,1]//momStratificationToTable[2,
	fancyPermutations->True]
\end{mmaCell}
	\begin{mmaCell}[verbatimenv=]{Output}
		\raisebox{-0.5\totalheight}{\includegraphics[width=0.9\linewidth]{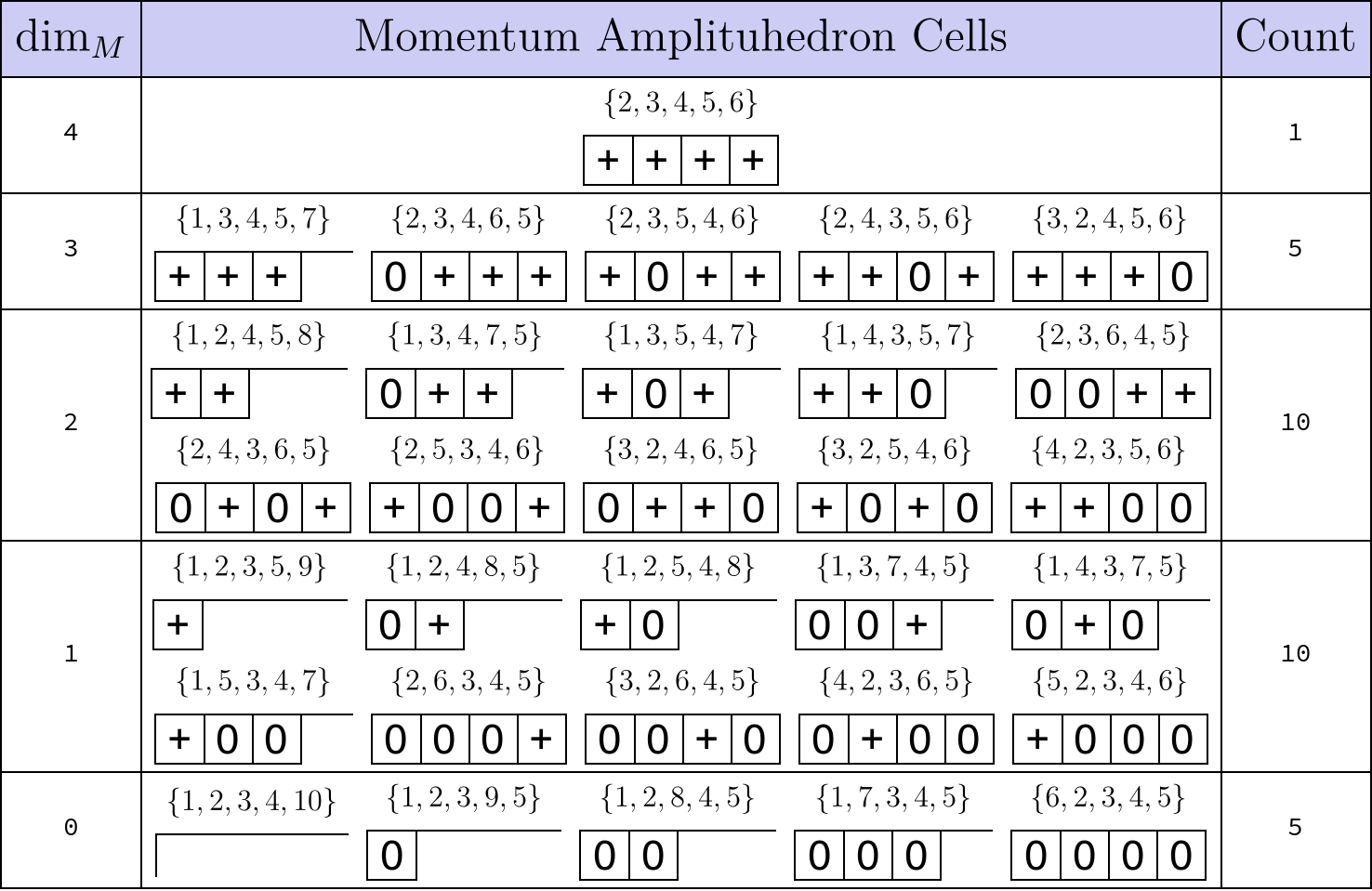}}
	\end{mmaCell}
}
		
		\mmaFunc{momStratificationToTable}{\mmaVar{m},\mmaVar{perm}} can be called using \\\mmaFunc{momStratificationToTable}{{\mmaVar{perm},m$\to$\mmaVar{m}}} or simply \\\mmaFunc{momStratificationToTable}{\mmaVar{perm}} where the default is {\tt m$\to$\mmaVar{m}}.
		
		\mmaFuncName{momStratificationToTable} additionally admits the following optional parameters:\\\\
		\mbox{\hspace{-1.5cm}\begin{tabular}{P{4cm}@{$\,$}c@{$\,$}P{2cm}@{$~$}c@{$~$}P{8.5cm}}
				\hline
				{Option}&&{Value}&&{Description}\\\hline\hline 
				{\tt showPermutations}&$\ast$&{\tt True}&{$\triangleright$}&displays permutation labels above the nodes of the Hasse diagram\\&&{\tt False}&{$\triangleright$}&leaves permutation labels above the nodes of the Hasse diagram off\\\hline
				{\tt fancyPermutations}&&{\tt True}&{$\triangleright$}&displays permutation labels in \LaTeX~font.\\&$\ast$&{\tt False}&{$\triangleright$}&leaves permutation labels unformatted.\\\hline
		\end{tabular}}
		
		For each optional parameter, we have indicated its default value by an asterix ``$\ast$''.
		
\item \mmaFuncDef{momIntervalToHasse}{\mmaVarDef{m},{\mmaVarDef{perm1},\mmaVarDef{perm2}}} returns a Hasse diagram for the poset of boundaries in {\tt momInterval[\mmaVar{m}][\mmaVar{perm1},\mmaVar{perm2}]}.
	
	\mmaFunc{momIntervalToHasse}{\mmaVar{m},{\mmaVar{perm1},\mmaVar{perm2}}} can be called using \\\mmaFunc{momIntervalToHasse}{{\mmaVar{perm1},\mmaVar{perm2},m$\to$\mmaVar{m}}} or simply \\\mmaFunc{momIntervalToHasse}{{\mmaVar{perm1},\mmaVar{perm2}}} where the default is {\tt m$\to$2}.
	
	\mmaFuncName{momIntervalToHasse} additionally admits the same optional parameters as {\tt momStratificationToHasse}.
	
\item \mmaFuncDef{momStratificationTo3D}{\mmaVarDef{m},\mmaVarDef{perm}} returns a graph depicting the combinatorial relation between zero-dimensional boundaries in \\{\tt momStratification[\mmaVar{m}][\mmaVar{perm}]} provided the cell in the momentum amplituhedron $\mathcal{M}^{(\mmaVar{m})}$ labelled by the decorated permutation \mmaVar{perm} has amplituhedron dimension less than or equal to $3$.
{\small
	\begin{mmaCell}[index=1,moredefined={topCell,momStratificationTo3D}]{Input}
momStratificationTo3D[2][topCell[4,2]]
	\end{mmaCell}
	\begin{mmaCell}[verbatimenv=]{Output}
		\raisebox{-0.5\totalheight}{\includegraphics[width=5cm]{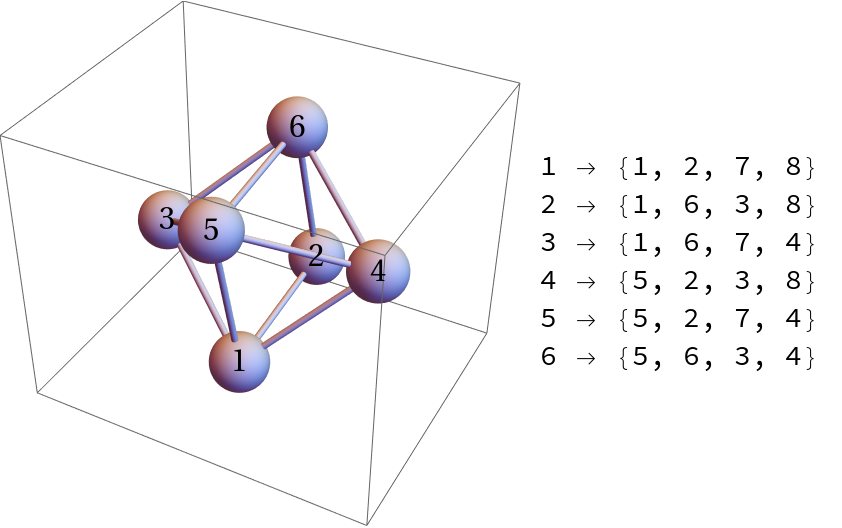}}
	\end{mmaCell}
}
	
	\mmaFunc{momStratificationTo3D}{\mmaVar{m},\mmaVar{perm}} can be called using \\\mmaFunc{momStratificationTo3D}{{\mmaVar{perm},m$\to$\mmaVar{m}}} or simply \\\mmaFunc{momStratificationTo3D}{\mmaVar{perm}} where the default is {\tt m$\to$2}.
	
	\mmaFuncName{momStratificationTo3D} additionally admits the following optional parameters:\\\\
	\mbox{\hspace{-1.5cm}\begin{tabular}{P{4cm}@{$\,$}c@{$\,$}P{2cm}@{$~$}c@{$~$}P{8.5cm}}
			\hline
			{Option}&&{Value}&&{Description}\\\hline\hline 
			{\tt showGraphics}&&{\tt True}&{$\triangleright$}&displays \Le-diagrams for each vertex in the graph\\&$\ast$&{\tt False}&{$\triangleright$}&does not display \Le-diagrams\\\hline
			{\tt showPermutations}&$\ast$&{\tt True}&{$\triangleright$}&displays permutation labels for each vertex in the graph\\&&{\tt False}&{$\triangleright$}&does not display permutation labels\\\hline
	\end{tabular}}
	
	For each optional parameter, we have indicated its default value by an asterix ``$\ast$''.
	
\item \mmaFuncDef{momFacetsToGraph}{\mmaVarDef{m},\mmaVarDef{perm}} returns a star-shaped graph of the facets (co-dimension $1$ boundaries) of the cell in the momentum amplituhedron $\mathcal{M}^{(\mmaVar{m})}$ labelled by the decorated permutation \mmaVar{perm} provided it has momentum amplituhedron dimension greater than or equal to 1.
	
{\small
	\begin{mmaCell}[index=1,moredefined={topCell,momFacetsToGraph}]{Input}
topCell[4,2]//momFacetsToGraph[2]
\end{mmaCell}
	\begin{mmaCell}[verbatimenv=]{Output}
		\raisebox{-0.5\totalheight}{\includegraphics[width=5cm]{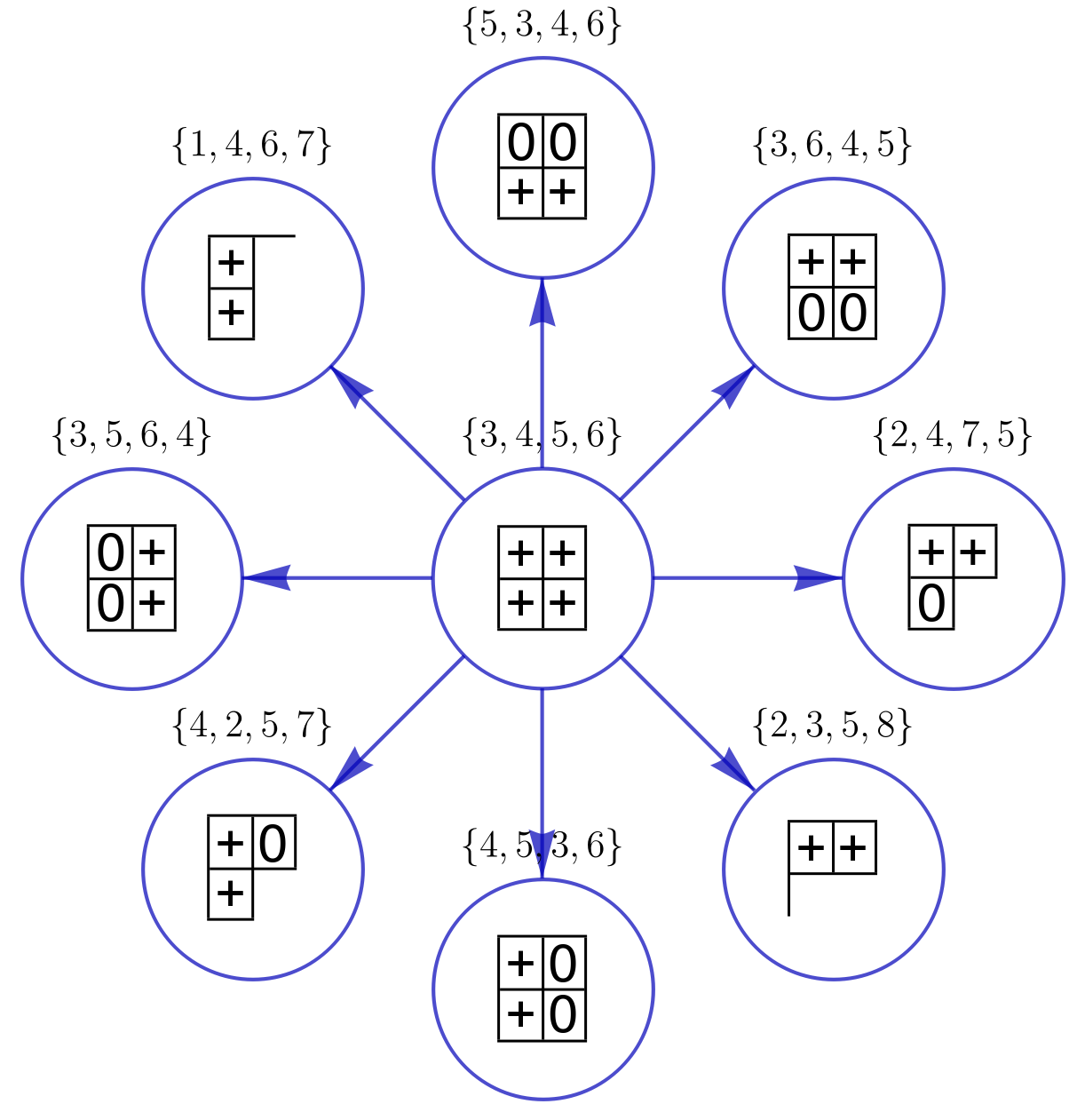}}
	\end{mmaCell}
}	
	
	\mmaFunc{momFacetsToGraph}{\mmaVar{m},\mmaVar{perm}} can be called using \\\mmaFunc{momFacetsToGraph}{{\mmaVar{perm},m$\to$\mmaVar{m}}} or simply \\\mmaFunc{momFacetsToGraph}{\mmaVar{perm}} where the default is {\tt m$\to$2}.
	
	\mmaFuncName{momFacetsToGraph} additionally admits the same optional parameters as {\tt momStratificationToHasse}.
\end{itemize}

The remaining list of functions exhibit analogous behaviour to the corresponding functions in Section \ref{sec:soft-amp}: 
\begin{itemize}
\item \mmaFuncDef{momDimension}{\mmaVarDef{m},\mmaVarDef{perm}}{see {\tt ampDimension}.}
\item \mmaFuncDef{momFacesQ}{\mmaVarDef{m},\mmaVarDef{perm}}{see {\tt ampFacesQ}.}
\item \mmaFuncDef{momFaces}{\mmaVarDef{m},\mmaVarDef{perm}}{see {\tt ampFaces}.}
\item \mmaFuncDef{momBoundaries}{\mmaVarDef{m},\mmaVarDef{perm}}{see {\tt ampBoundaries}.}
\item \mmaFuncDef{momStratification}{\mmaVarDef{m},\mmaVarDef{perm}}{see {\tt ampStratification}.}
\item \mmaFuncDef{momInverseStratification}{\mmaVarDef{m},\mmaVarDef{perm}}{see \\{\tt ampInverseStratification}.}
\item \mmaFuncDef{momBoundary}{\mmaVarDef{m},\mmaVarDef{perm}}{see {\tt ampBoundary}.}
\item \mmaFuncDef{momInverseBoundary}{\mmaVarDef{m},\mmaVarDef{perm}}{see {\tt ampInverseBoundary}.}
\item \mmaFuncDef{momInterval}{\mmaVarDef{m},{\mmaVarDef{perm1},\mmaVarDef{perm1}}}{see {\tt ampInterval}.}
\item \mmaFuncDef{momPoset}{\mmaVarDef{m},\mmaVarDef{perm}}{see {\tt ampPoset}.}
\item \mmaFuncDef{momPoset}{\mmaVarDef{m},{\mmaVarDef{perm1},\mmaVarDef{perm1}}}{see {\tt ampPoset}.}
\end{itemize}

\subsection{\emph{hyp} -- hypersimplex boundaries}\label{sec:soft-hyp}
\begin{itemize}
	
	\item \mmaFuncDef{hypBasis}{\mmaVarDef{perm}} returns a list of all $k$-length subsets of $[n]=\{1,\dots,n\}$ (where\footnote{{\tt permK} is a function in {\tt positroids} \cite{Bourjaily:2012gy}.} $k=\text{\tt permK[\mmaVar{perm}]}$ and $n=\text{\tt Length[\mmaVar{perm}]}$) which label the non-zero ordered maximal $k\times k$ minors of {\tt permToMat[\mmaVar{perm}]}.
{\small
	\begin{mmaCell}[index=1,moredefined={topCell,hypBasis}]{Input}
topCell[4,2]//hypBasis
	\end{mmaCell}
	\begin{mmaCell}{Output}
\{\{1,2\},\{1,3\},\{1,4\},\{2,3\},\{2,4\},\{3,4\}\}
	\end{mmaCell}
}	
	
\item \mmaFuncDef{hypBases}{{\mmaVarDef{n},\mmaVarDef{k}}} returns the output {\tt hypBasis} evaluated for \emph{all} cells in the positroid stratification of the positive Grassmannian $G_+(\mmaVar{k},\mmaVar{n})$.
{\small
	\begin{mmaCell}[index=1,moredefined={hypBases}]{Input}
hypBases[3,1]
	\end{mmaCell}
	\begin{mmaCell}{Output}
\{\{\{1\},\{2\},\{3\}\},\{\{2\},\{3\}\},\{\{1\},\{2\}\},\{\{1\},\{3\}\},\newline\{\{3\}\},\{\{2\}\},\{\{1\}\}\}
	\end{mmaCell}
}
		
		The first-time behaviour of this function depends on whether or not its output already exists on file in the {\tt Data/} folder. If this is the case, then this output is simply retrieved and stored in the kernel's memory. Otherwise the output is calculated directly and stored in the kernel's memory, and if {\tt \$cache=True} then the output is also then stored to file. 
		
		A message is displayed the first time the function is executed analogous to the messages displayed by {\tt ampDimension} detailing its first-time behaviour.
\end{itemize}

The remaining list of functions exhibit analogous behaviour to the corresponding functions in Section \ref{sec:soft-mom}: 

\begin{itemize}
\item \mmaFuncDef{hypDimension}{\mmaVarDef{perm}}{see {\tt momDimension}.}
\item \mmaFuncDef{hypFaceQ}{\mmaVarDef{perm}}{see {\tt momFaceQ}.}
\item \mmaFuncDef{hypFaces}{\mmaVarDef{perm}}{see {\tt momFaces}.}
\item \mmaFuncDef{hypBoundaries}{\mmaVarDef{perm}}{see {\tt momBoundaries}.}
\item \mmaFuncDef{hypStratification}{\mmaVarDef{perm}}{see {\tt momStratification}.}
\item \mmaFuncDef{hypInverseStratification}{\mmaVarDef{perm}}{see \\{\tt momInverseStratification}.}
\item \mmaFuncDef{hypBoundary}{\mmaVarDef{perm}}{see {\tt momBoundary}.}
\item \mmaFuncDef{hypInverseBoundary}{\mmaVarDef{perm}}{see {\tt momInverseBoundary}.}
\item \mmaFuncDef{hypInterval}{{\mmaVarDef{perm1},\mmaVarDef{perm2}}}{see {\tt momInterval}.}
\item \mmaFuncDef{hypPoset}{\mmaVarDef{perm}}{see {\tt momPoset}.}
\item \mmaFuncDef{hypPoset}{{\mmaVarDef{perm1},\mmaVarDef{perm2}}}{see {\tt momPoset}.}
\item \mmaFuncDef{hypStratificationToHasse}{\mmaVarDef{perm}}{see {\tt momStratificationToHasse}.}
\item \mmaFuncDef{hypStratificationToTable}{\mmaVarDef{perm}}{see {\tt momStratificationToTable}.}
\item \mmaFuncDef{hypIntervalToHasse}{\mmaVarDef{perm1},\mmaVarDef{perm2}}{see {\tt momIntervalToHasse}.}
\item \mmaFuncDef{hypStratificationTo3D}{\mmaVarDef{perm}}{see {\tt momStratificationTo3D}.}
\item \mmaFuncDef{hypFacetsToGraph}{\mmaVarDef{perm}}{see {\tt momFacetsToGraph}.}
\end{itemize}

%
%
%
%

\section{An Illustrative Example}
\label{sec:ex}

In many cases the amplituhedron recovers familiar objects: when $n=k+m$, i.e.\ when $Z$ is a square matrix, then $\mathcal{A}_{k+m,k}^{(m)}$ is just the positive Grassmannian $G_+(k,k+m)$; while for $k=1$ it is a cyclic polytope \cite{STURMFELS1988275}. Moreover, when $k=n-m-1$ then, using the parity duality explored in \cite{Galashin:2018fri,LPW}, one can show that amplituhedron shares many properties with cyclic polytopes. For this reason, for $m=2$, the first non-trivial example is $\mathcal{A}_{6,2}^{(2)}$. In this section we apply our package to study its properties: we find its boundary stratification and its poset, and show how to use our package to visualize this data for further studies. 

\subsection{The boundary stratification of the top cell}

The stratification of boundaries in amplituhedron $\mathcal{A}^{(2)}_{6,2}$ can be calculated using the function {\tt ampStratification}:
{\small
\begin{mmaCell}[index=1,moredefined={topCell,ampStratification,Counts,ampDimension,Length}]{Input}
topCell[6,2]
Counts[ampDimension/@ampStratification[\%]]
\end{mmaCell}
\begin{mmaCell}{Output}
\{3,4,5,6,7,8\}
\end{mmaCell}
\begin{mmaCell}{Message}
ampStratification: called for the first time with n=6, k=2, m=2; 
	loading definition from "<notebook directory>/Data/
	ampStratification-6-2-2.m".
\end{mmaCell}
\begin{mmaCell}{Message}
ampDimension: called for the first time with n=6, k=2, m=2; loading 
	definition from "<notebook directory>/Data/
	ampDimension-6-2-2.m".
\end{mmaCell}
\begin{mmaCell}{Output}
<|4->1,3->6,2->21,1->30,0->15|>
\end{mmaCell}
}
\noindent The amplituhedron $\mathcal{A}^{(2)}_{6,2}$ is the image of the top-cell for the positive Grassmannian $G_{+}(2,6)$, labelled by the decorated permutation {\tt \{3,4,5,6,7,8\}}. The above output gives us the number of boundaries of each dimension in the amplituhedron. This information is usually encoded in the so-called $f$-vector, which counts the number of boundaries of each dimension, and this can be computed via\footnote{The $f$-vector is defined as the tuple $f=(f_{-1},f_0,f_1,\ldots f_d)$ where $f_i$ counts the number of boundaries of dimension $i$ (for $1\le i\le d$) and we trivially set $f_{-1}=1$.}
{\small
\begin{mmaCell}[moredefined={List,Reverse,Join}]{Input}
Join[\{1\},Reverse[List@@\%]]
\end{mmaCell}
\begin{mmaCell}{Output}
\{1,15,30,21,6,1\}
\end{mmaCell}
}
\noindent and from the $f$-vector one can show that the Euler characteristic for this geometry is 0
{\small
\begin{mmaCell}[moredefined={Range,Length}]{Input}
\%.(-1)\^{Range[Length[\%]]}
\end{mmaCell}
\begin{mmaCell}{Output}
0
\end{mmaCell}
}
\noindent as expected \cite{Lukowski:2019kqi}. 

More detailed information about the boundary stratification of amplituhedron can be obtained from the function {\tt ampStratificationToTable} and is given in \tbl{tbl:ampStrat-polygon},
\begin{table}
{\small
	\begin{mmaCell}[moredefined={topCell,ampStratificationToTable}]{Input}
topCell[6,2]//ampStratificationToTable[2,
	fancyPermutations->True]
	\end{mmaCell}
	\begin{mmaCell}[verbatimenv=]{Output}
\raisebox{-0.5\totalheight}{\includegraphics[height=0.9\textheight]{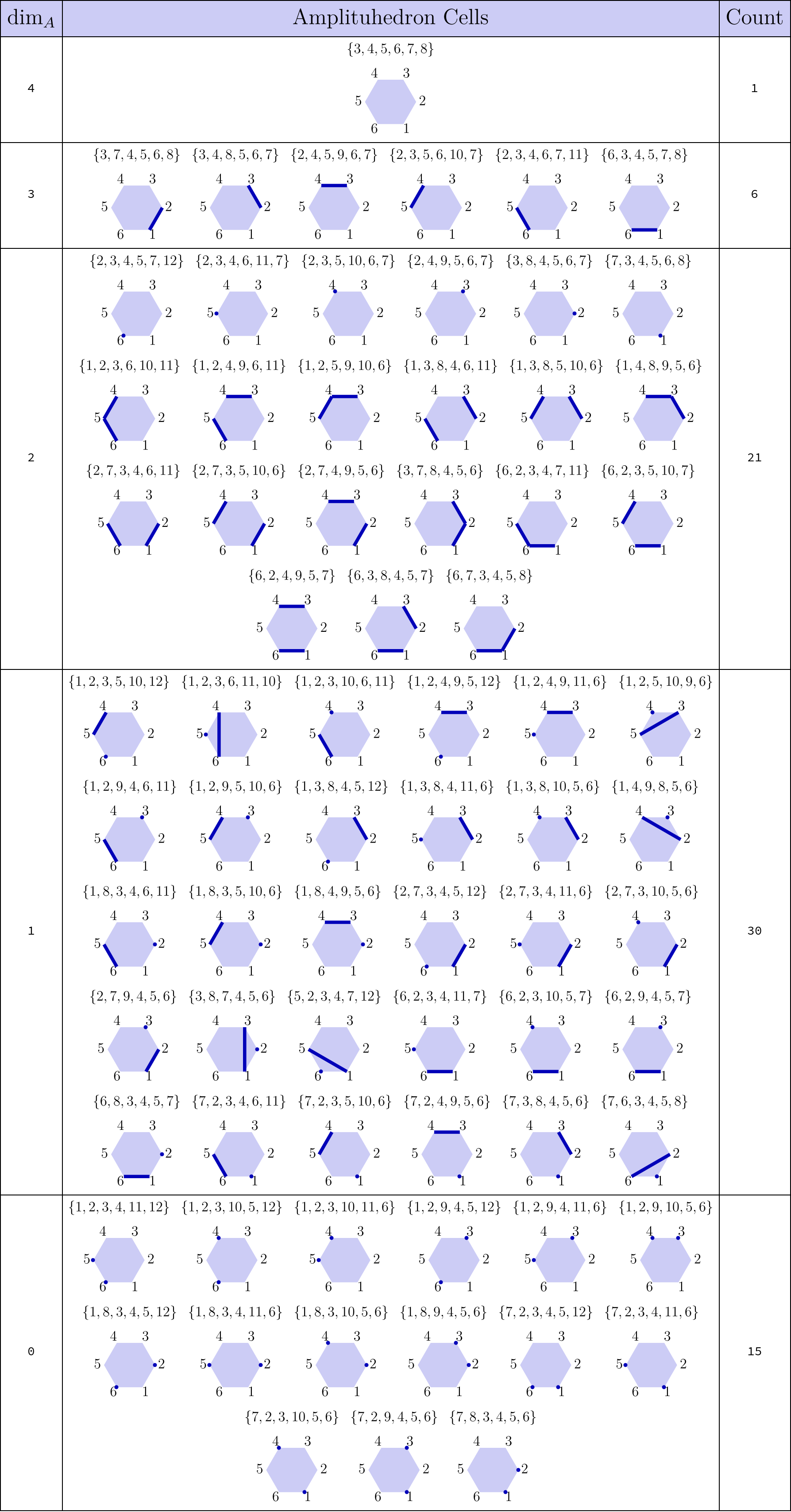}}
	\end{mmaCell}
}
\vspace{-0.5cm}
\caption{Boundary stratification of the amplituhedron $\mathcal{A}^{(2)}_{6,2}$ with labels from \cite{Lukowski:2019kqi} produced by {\tt ampStratificationToTable}.}
\label{tbl:ampStrat-polygon}
\end{table}
where each decorated permutation is labelled by a hexagonal graphic (the details of this graphical enumeration are presented in \cite{Lukowski:2019kqi}). Alternatively, one can choose to label each decorated permutation instead by \Le-diagrams, and this is shown in \tbl{tbl:ampStrat-lediagram}.
\begin{table}
{\small
	\begin{mmaCell}[moredefined={topCell,ampStratificationToTable}]{Input}
topCell[6,2]//ampStratificationToTable[2,
	fancyPermutations->True,
	graphicType->"lediagram"]
	\end{mmaCell}
	\begin{mmaCell}[verbatimenv=]{Output}
\raisebox{-0.5\totalheight}{\includegraphics[height=0.85\textheight]{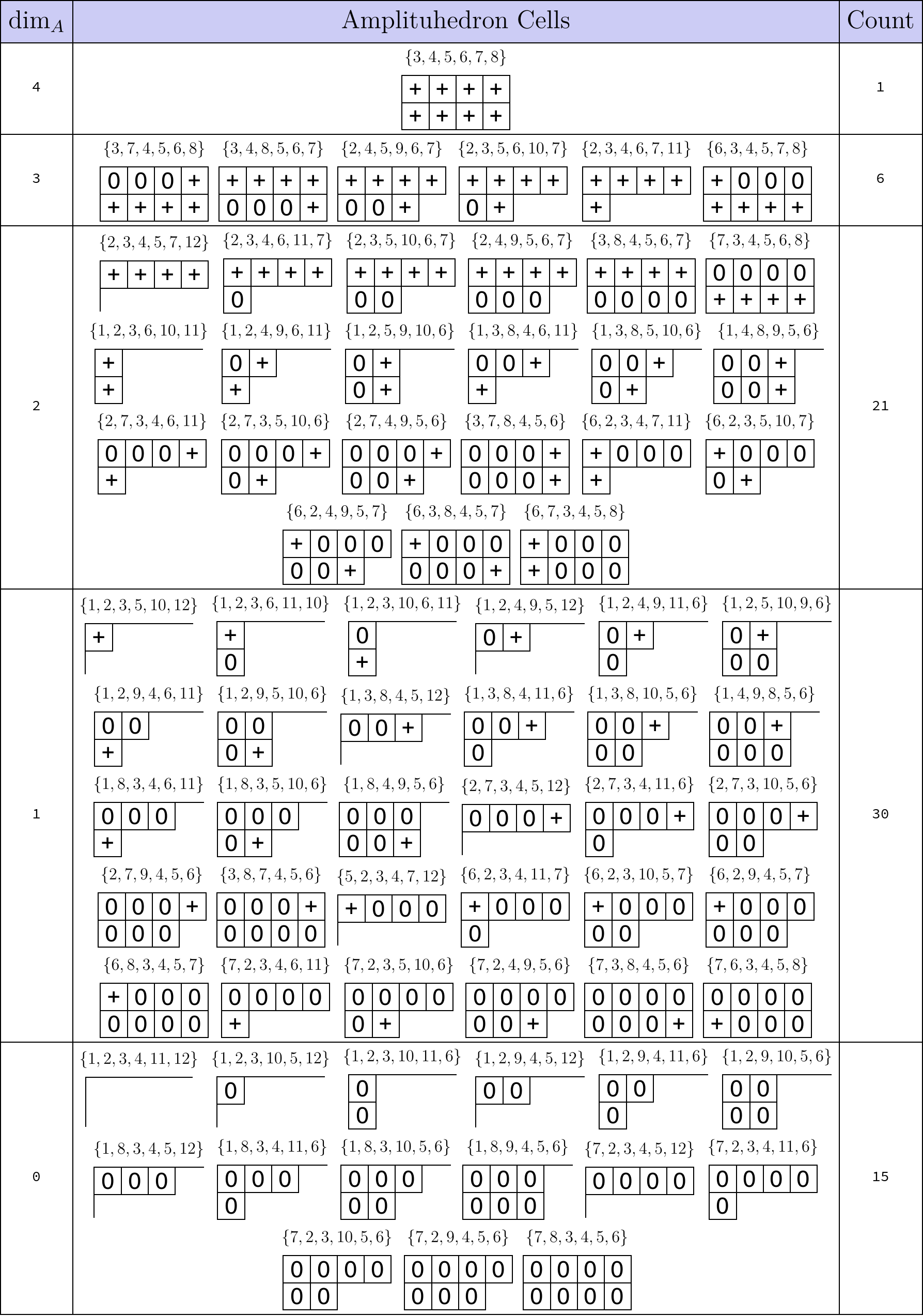}}
	\end{mmaCell}
}
\vspace{-0.5cm}
\caption{Boundary stratification of the amplituhedron $\mathcal{A}^{(2)}_{6,2}$ with labels given by \Le diagrams produced by {\tt ampStratificationToTable}.}
\label{tbl:ampStrat-lediagram}
\end{table}

From the above two tables we can read off the following information:
\begin{itemize}
	\item the top-cell in the amplituhedron is $4$-dimensional.
	\item it has $6$ facets (or $3$-dimensional cells), each labelled uniquely by one of the $6$ edges of the regular hexagon. All of these boundaries are the image of $5$-dimensional positroid cells in the positive Grassmannian; their positroid dimensions can be read off from the \Le-diagrams by counting the number of plus ``$+$'' symbols.
	\item there are $21$ co-dimension $2$ boundaries (ridges) which come in two types: 
	\begin{enumerate}
		\item $6$ ridges labelled by the $6$ vertices of the regular hexagon. Each of these boundaries is the image of a $4$-dimensional positroid cell.
		\item $15$ ridges labelled by all combinations of two edges of the regular hexagon, and these boundaries all have $2$-dimensional preimages under the amplituhedron map.
	\end{enumerate}
	\item the total number of edges is $30$. These are labelled by all combinations of one of the $6$ vertices of the regular hexagon together with one of the $5$ edges of a regular pentagon. From the \Le-diagrams we read of the dimensions of their preimages to all be $1$.
	\item finally there are $15$ vertices in the amplituhedron labelled by all combinations of two vertices of the regular hexagon.
\end{itemize}

\subsection{Amplituhedron Facets}
The facets of the amplituhedron $\mathcal{A}^{(2)}_{6,2}$ can be found in at least the following ways:
\begin{itemize}
	\item as boundaries of the top cell in the amplituhedron
{\small
\begin{mmaCell}[moredefined={topCell,ampBoundary}]{Input}
ampBoundary[topCell[6,2]]
\end{mmaCell}
}
\item as all co-dimension $1$ boundaries of the top cell in the amplituhedron
{\small
	\begin{mmaCell}[moredefined={topCell,ampBoundaries}]{Input}
ampBoundaries[1][topCell[6,2]]
	\end{mmaCell}
}
\end{itemize}

Given that we already have the boundary stratification of the amplituhedron $\mathcal{A}^{(2)}_{6,2}$, we can now investigate the relation between a given stratum and its boundaries. To this end, let us consider one of the amplituhedron facets, e.g.\ the one labelled by {\tt \{6,3,4,5,7,8\}}.
{\small
\begin{mmaCell}[moredefined={topCell,ampFacetsToGraph}]{Input}
ampFacetsToGraph[\{6,3,4,5,7,8\}]
\end{mmaCell}
\begin{mmaCell}[verbatimenv=]{Output}
\raisebox{-0.5\totalheight}{\includegraphics[width=6cm]{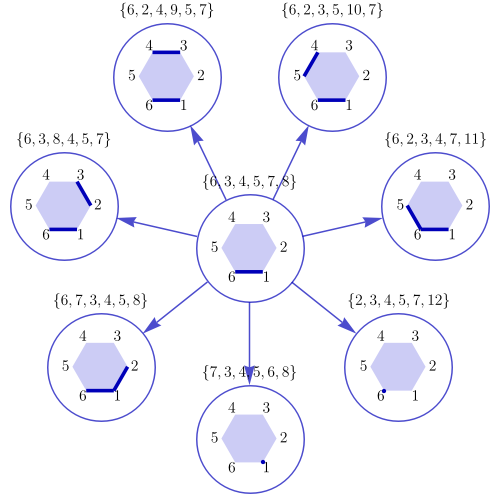}}
\end{mmaCell}
}

Being each $3$-dimensional, the combinatorial structure of each facet can be depicted graphically using {\tt ampStratificationTo3D}:
{\small
	\begin{mmaCell}[moredefined={topCell,ampBoundary,ampStratificationTo3D}]{Input}
ampStratificationTo3D[\{6,3,4,5,7,8\}]
	\end{mmaCell}
	\begin{mmaCell}[verbatimenv=]{Output}
\raisebox{-0.5\totalheight}{\includegraphics[width=8cm]{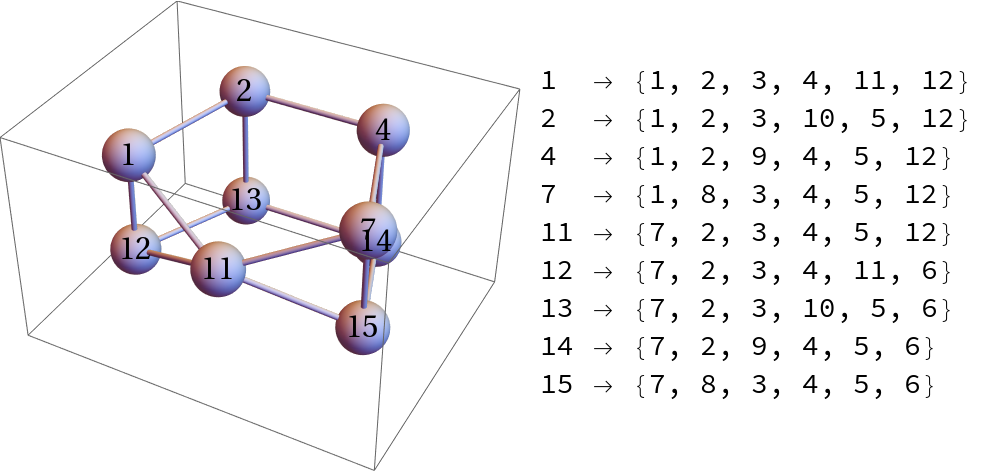}}
	\end{mmaCell}
}
\noindent where the labelled nodes denote the zero-dimensional vertices of the facet. All six facets of the amplituhedron $\mathcal{A}^{(2)}_{6,2}$ share the same combinatorial structure: $9$ vertices, $14$ edges, and $7$ faces consisting of $2$ triangles, $3$ quadrilaterals and $2$ pentagons.

\begin{figure}
\includegraphics[width=\linewidth]{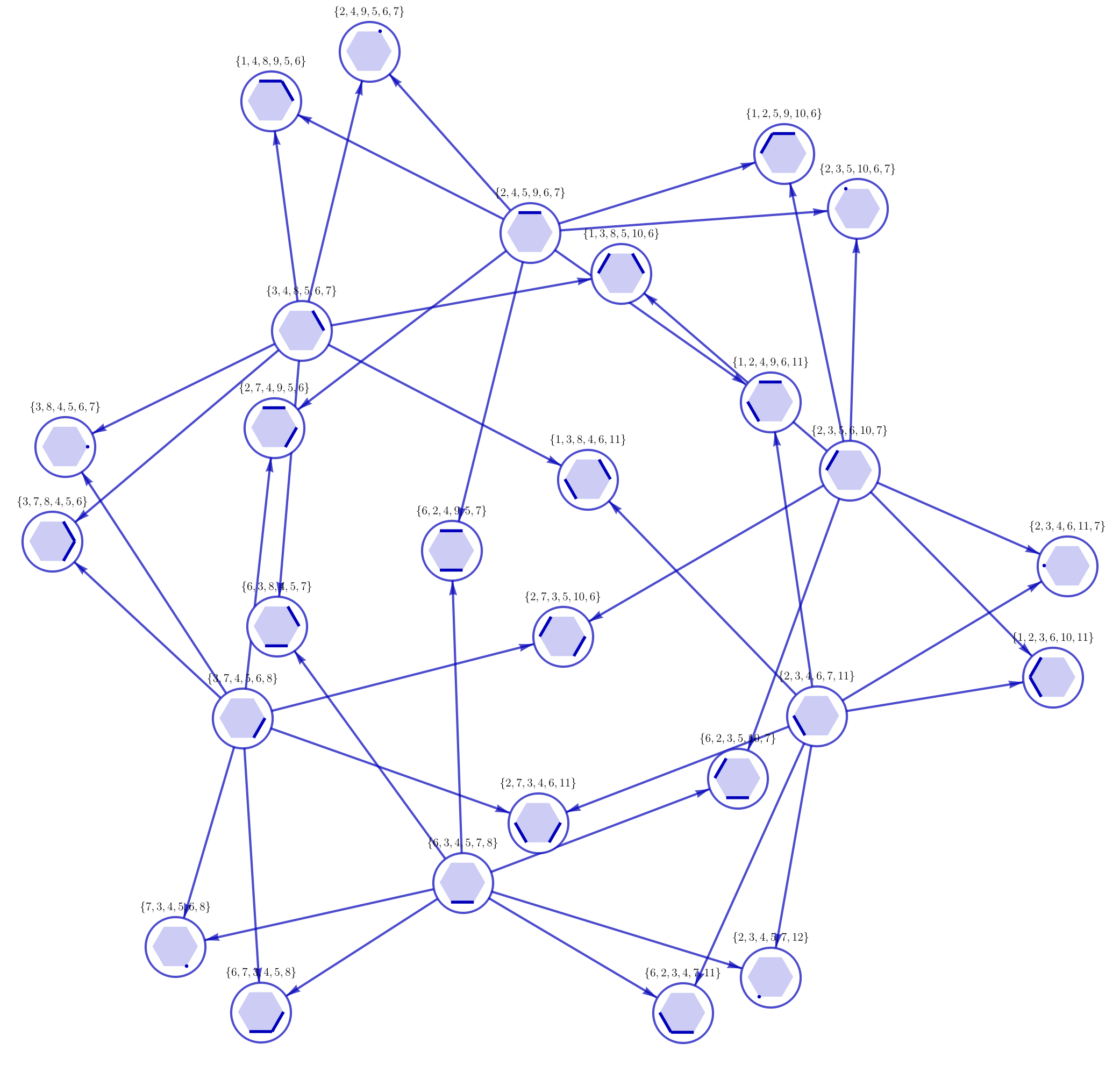}
\caption{The intersection of facets of the amplituhedron $\mathcal{A}^{(2)}_{6,2}$ at ridges.}
\label{fig:ex-amp-facets-ridges-6-2}
\end{figure}

\begin{figure}
\begin{subfigure}[b]{.489\textwidth}
	\includegraphics[width=0.9\linewidth]{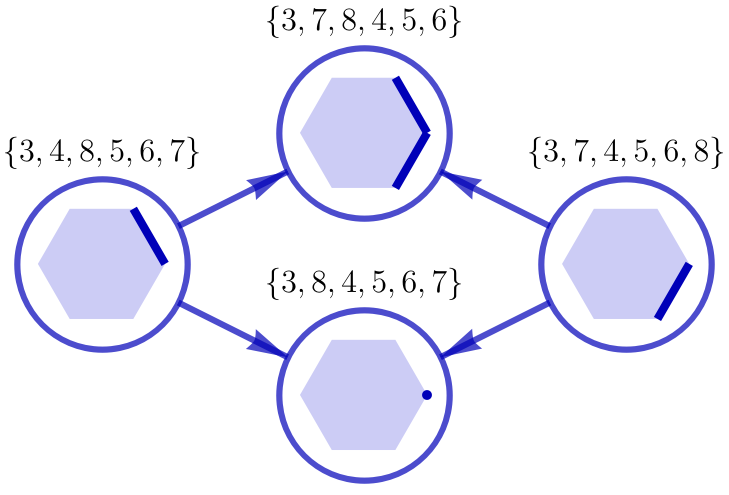}
	\caption{Facets sharing two ridges.}
	\label{fig:amp-facets-ridges-polygons-adjacent}
\end{subfigure}
\hfill
\begin{subfigure}[b]{.489\textwidth}
	\raisebox{1.11cm}{\includegraphics[width=0.9\linewidth]{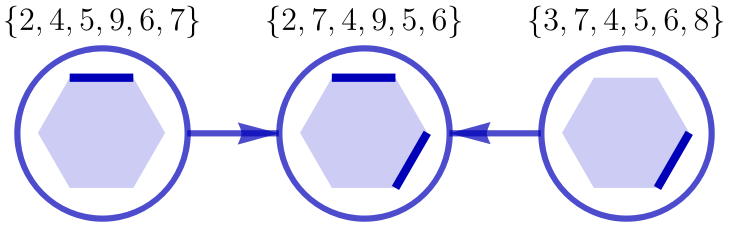}}
	\caption{Facets sharing only one ridge.}
	\label{fig:amp-facets-ridges-polygons-non}
\end{subfigure}
\caption{Graphical depiction of ridges shared between pairs of facets in the amplituhedron $\mathcal{A}^{(2)}_{6,2}$. Each cell is labelled by a graphic produced by {\tt ampPermToGraphic[2,showLabels->False]}.}
\label{fig:amp-facets-ridges-polygons}
\end{figure}

\begin{figure}
	\begin{subfigure}{\textwidth}
		\includegraphics[width=\linewidth]{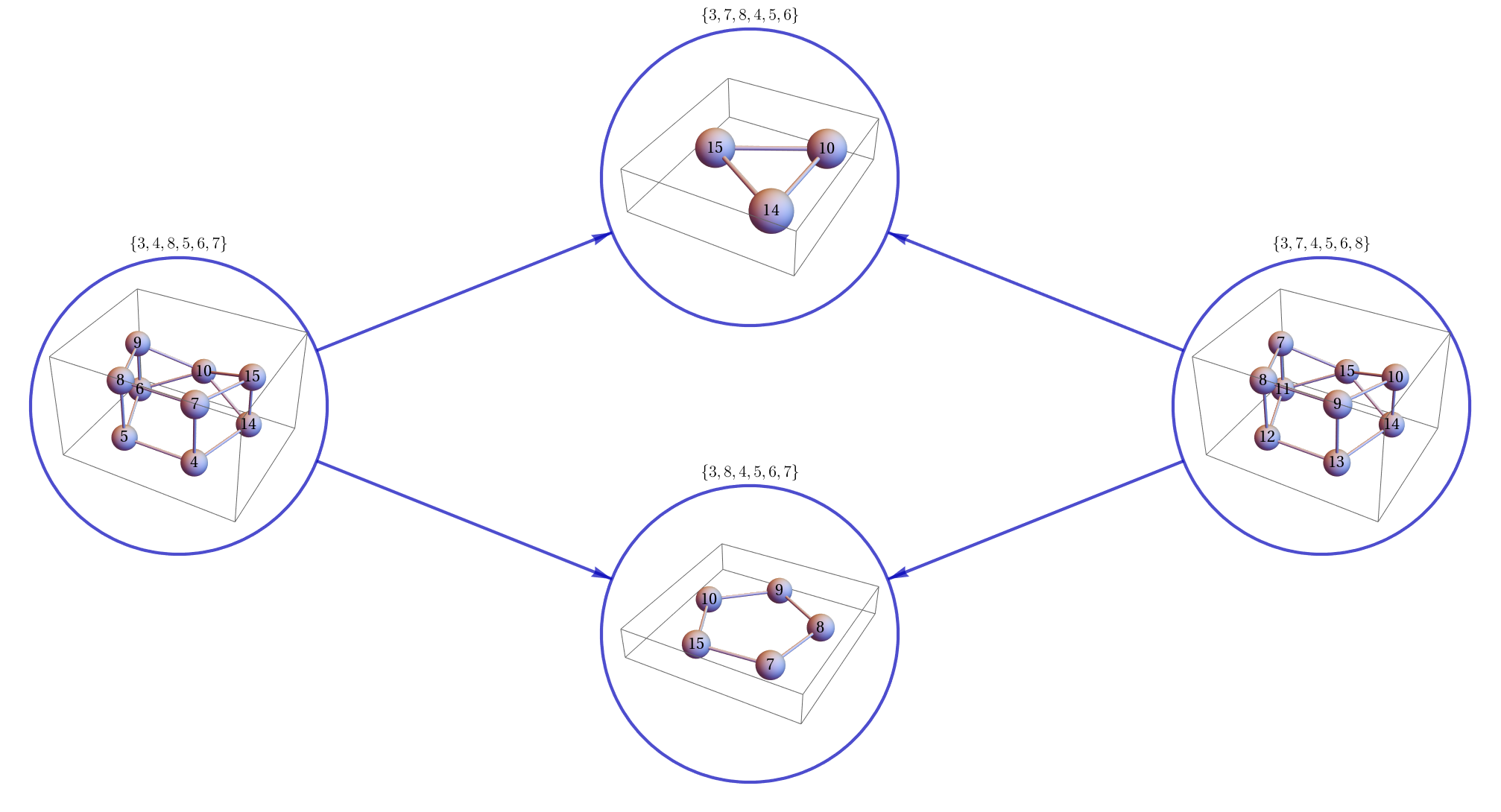}
		\caption{Facets sharing two ridges.}
		\label{fig:amp-facets-ridges-graphs-adjacent}
	\end{subfigure}
	\begin{subfigure}{\textwidth}
		\includegraphics[width=\linewidth]{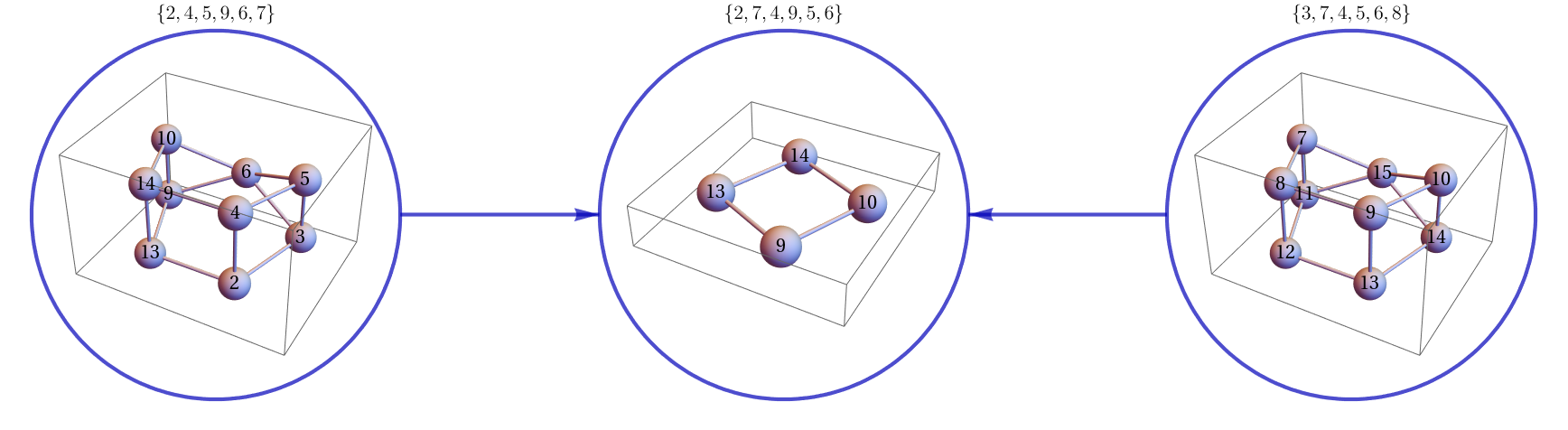}
		\caption{Facets sharing only one ridge.}
		\label{fig:amp-facets-ridges-graphs-non}
	\end{subfigure}
	\caption{Graphical depiction of ridges shared between pairs of facets in the amplituhedron $\mathcal{A}^{(2)}_{6,2}$. Each cell is labelled by a combinatorial graph produced by {\tt ampStratificationToGraph[2,showPermutations->False]}.}
	\label{fig:amp-facets-ridges-graphs}
\end{figure}

The relationship between facets, specifically how they glue together at ridges, is depicted in \fig{fig:ex-amp-facets-ridges-6-2}. Notice that there are two types of connections between facets which are illustrated in  \fig{fig:amp-facets-ridges-polygons} and \fig {fig:amp-facets-ridges-graphs}. For pairs of facets defined by the conditions
\begin{equation}
\langle Y\,i\,i+1\rangle=0\qquad \langle Y\,i+1\,i+2\rangle=0,
\end{equation}
they share two ridges, while for other pairs of facets they share only one ridge.

\section{Conclusions}
\label{Sec:Conclusions}
In this paper, we introduced a new \Mathematica package \begin{center}``\texttt{amplituhedronBoundaries}''\end{center} to help explore positive geometries relevant for scattering amplitudes. The main aim of this package is to provide tools for the study of boundaries and relations between them for three positive geometries: the amplituhedron, the momentum amplituhedron and the hypersimplex. In particular, it allows us to reproduce all the results on boundary stratifications for $\mathcal{A}_{n,k}^{(2)}$, presented in \cite{Lukowski:2019kqi}. It also provides the complete stratification of the hypersimplex $\Delta_{k,n}$ in terms of images of positroid cells in $G_+(k,n)$ through a moment map. Equivalently, it computes the boundary stratification of the momentum amplituhedron $\mathcal{M}_{n,k}^{(2)}$. Its functionality extends beyond the case of $m=2$ and includes many functions which provide preliminary results for stratifications in the physical, $m=4$, case.

\section*{Acknowledgements}
\label{Sec:Acknowledgements}
We are very grateful to Jacob Bourjaily for discussions on the implementation of some of the package functions. TL would like to thank Livia Ferro, Matteo Parisi, Andrea Orta, David Damgaard, Anastasia Volovich, Marcus Spradlin and Lauren Williams for collaborations which motivated the development of this \Mathematica package.





\bibliographystyle{utphys}
\bibliography{amplituhedronBoundaries}

\end{document}